\documentclass[useAMS]{mn2e}
\usepackage{amsmath}
\usepackage{textcomp}
\usepackage{amssymb}
\usepackage[pdftex]{graphicx}
\usepackage{amssymb}

\usepackage[margin=.8in, paperwidth=8.5in, paperheight=11in]{geometry}
\usepackage{multicol}
\usepackage{epsfig}
\usepackage{graphicx}
\usepackage{dblfloatfix}

\usepackage{accents}
\newcommand*{\dt}[1]{%
  \accentset{\mbox{\Large\bfseries .}}{#1}}

\begin{document}

\title{Asymmetric MHD Outflows/Jets from Accreting T Tauri Stars}
\author[S. Dyda et al.]
{\parbox{\textwidth}{S.~Dyda$^{1}$,
R.V.E.~Lovelace$^{2}$,
G.V.~Ustyugova$^{3}$,
P.S. Lii$^{2}$,
M.M.~Romanova$^{2}$,\\
 \& A.V.~Koldoba$^{4,5}$}\vspace{0.3cm}\\
 $^{1}$Department of Physics, Cornell University, Ithaca, NY 14853;  email: sd449@cornell.edu\\
 $^{2}$Department of Astronomy, Cornell University, Ithaca, NY 14853\\
$^{3}$Keldysh Institute for Applied Mathematics, Moscow, Russia\\
$^{4}$Moscow Institute of Physics \& Technology, Dolgoprudnyy, Moscow Region, Russia\\
$^{5}$Institute of Computer Aided Design RAS, Moscow, 123056, Russia\\
}

\date{\today}
\pagerange{\pageref{firstpage}--\pageref{lastpage}}
\pubyear{2015}

\label{firstpage}

\maketitle

\begin{abstract}

Observations of jets from young stellar objects reveal the asymmetric outflows
from some sources. A large set of $2.5$D MHD simulations has been carried out for  axisymmetric 
viscous/diffusive disc accretion to rotating magnetized stars for the 
purpose of assessing the conditions where the outflows or jets
are asymmetric relative to the equatorial plane.     
     We consider initial magnetic fields that
are symmetric about the equatorial plane and consist of a radially 
distributed  field threading the disc (disc-field) and a stellar dipole field.
 ({\bf 1}). For pure disc-fields the symmetry or asymmetry of the outflows is affected by  the midplane plasma $\beta$ of the disc 
  (where $\beta$ is the ratio of the plasma pressure to
 the magnetic pressure).
    For the low density discs with small plasma $\beta$ values, outflows are observed to be symmetric to
 within $10\%$ over timescales of hundreds of inner disc orbits.
      For the denser higher $\beta$  discs, the coupling of the upper
 and lower coronal plasmas is broken, and quasi-periodic field motion in
the two hemispheres becomes different.   This asymmetry leads to
 asymmetric episodic outflows. 
({\bf 2.}) Accreting stars with a  stellar dipole field and
no disc-field exhibit episodic, two component outflows - a magnetospheric wind and an inner disc wind
from somewhat larger radial distances.
Both are characterized by similar velocity profiles but the magnetospheric wind has densities $\gtrsim 10$ times that of the disc wind. 
({\bf 3}.)Adding a disc-field parallel to the stellar dipole field acts
 to enhance the magnetospheric winds but suppress the disc wind.  
({\bf 4}.)  In contrast, adding a disc-field which is anti-parallel to the stellar dipole field in the disc acts
to suppress the magnetospheric and disc winds.
Our simulations reproduce some key features of observations of asymmetric outflows of T Tauri stars.

\end{abstract}

\begin{keywords} accretion,  accretion discs -- MHD -- black hole physics, magnetic fields, jets, stars: winds, outflows
\end{keywords}

\section{Introduction}

   Astrophysical jets are observed from many disc accreting objects
ranging from Young Stellar Objects (YSOs)  to systems with white dwarfs, neutron stars and black holes (BHs) (e.g.,  Livio 1997).
The jets transport angular
momentum away from the disc and thereby facilitate accretion
to the central object.   A large-scale magnetic field threading the
disc is thought to have a key role in launching and collimating 
the jets (e.g., Lovelace et al. 2014).

   There is clear evidence mainly from {\it Hubble Space Telescope} (HST)
 observations that YSOs often show  an asymmetry between   the jet and the counter jet. In HH30 the jet has more structure than the counter jet in the form of knots, suggesting different  mass densities or different acceleration mechanisms (Mundt et al. 1990;  Ray et al. 1996). The mass flux of the jet is approximately constant along its entire length whereas the counterjet's mass flux varies by a factor of $2$ along its length (Bacciotti et al. 1999). RW Aur A exhibits a jet with velocity approximately twice that of its counter jet and a mass flux $2-3$ times as large (Woitas et al. 2002, Hartigan \& Hillenbrand 2009, Melnikov et al. 2009). Likewise LkH$\alpha$ 233 exhibits a jet with a knotted structure, whereas its counter jet has no knots but has similar velocity (Perrin \& Graham 2007). FS Tau B has been observed to have a jet with a higher velocity but a lower density than its counter jet (Liu et al. 2012). Rotation measurements of T Tauri jets find evidence of jets counter-rotating with respect to the accretion disc or of the opposite rotation of jets in the northern and southern hemispheres (Sauty et al. 2012).  Studies have shown that the jet/counter jet structure are different, with radial velocity differences of $\sim 10-25$ km/s (Coffey et al. 2004;  Ray et al. 2007 for a review).
  One of the few exceptions is HH 212 which exhibits nearly symmetric jet/counterjet (Zinnecker 1998).  Ellerbroek et al. (2013) found HH 1042 to have jets with similar mean velocities but the knotted structure and velocity spread suggests an asymmetric launching mechanism.  
     Whelan et al. (2014) have studied the jet/counter jets mass outflow rates and stellar accretion rates in ESO-H$\alpha$ Par-Lup 3-4 to constrain lauching mechanisms. Though most observations show some sort of asymmetry in the outflows, this may be a result of asymmetries in the ambient medium and not the launching mechanism (Carratti o Garatti 2013, White 2014).     

    Relatively little theoretical and computational work has been directed
at understanding the asymmetry of the YSO jets.   
    A theoretical study by Wang et al. (1992) considered disc magnetic fields
composed of dipole and quadrupole components and 
found that the magnetically driven jets could in some cases be highly asymmetric.    Chagelishvili et al. (1996) investigated how a quadrupolar disc dynamo interacting with a stellar dipole could give rise to one sided jets. 
    Lovelace et al. (2010) carried out axisymmetric MHD simulations
 of disc accretion to stars with magnetic field consisting of different
 combinations of aligned dipole
 and quadrupole field components and without the assumption of symmetry about the  equatorial plane.   
      The simulations showed
that for conditions where there is a significant quadrupole component in addition to the dipole component, a dominantly one-sided conical wind tends to form on the side of the equatorial plane with the larger value of the intrinsic axial magnetic field.  
  For cases where the quadrupole component is absent or very small, dominantly one-sided outflows also form, but the direction of the flow Ôflip-flopsÕ between upward and downward on a time-scale of  $\sim 30$ d for a protostar.
    von Rekowski  (2003) and von Rekowski \& Brandenburg  (2006) investigated outflows generated by time dependent stellar and disc dynamos.
    Matsakos et al. (2012) investigated how field alignment between the disc and star and pressure distribution in the ambient cloud caused velocity asymmetries in protostellar jets.
      Fendt and Sheikhnezami  (2013)  investigated how thermal and
 density asymmetries in the disc can lead to asymmetries in the outflow.
       Velazquez (2014) performed 3D hydrodynamical simulations to study the case of asymmetric jet launching from protoplanetary discs that are preferentially fed matter from one side of the disc by a binary companion. 
         Stute et al. (2014) showed that relaxing axisymmetry has little effect on MHD driven winds.    

    Classical T Tauri stars (CTTS) commonly show evidence of a stellar dipole and higher order multipoles with field strengths $\sim 1-2$ kG at the 
stellar surface (Jardine, Collier Cameron \& Donati 2002; 
Donati et al. 2007, 2008).
Furthermore, it is known from MHD simulations that
any  large-scale magnetic field threading the star's accretion disc tends to
be advected inward due partly to viscous transport in the disc and partly due to angular momentum loss to outflows (e.g., Dyda et al. 2012).
    A  large-scale magnetic field is thought
to be responsible for driving and collimating the jets from
young stellar objects.
     The origin of the required magnetic fields is not established,
though a strong possibility is that dynamo processes in the 
rotating, convective stars.  
    For example,  the Zeeman signature in GQ Lup (Donati et al. 2012) and DN Tau (Donati et al. 2013) suggests that the magnetic fields of T Tauri stars are generated by non-stationary dynamos. 
    Observations of disc-accreting stars show strong
correlation between the magnitude of the star's dipole magnetic field
and the rotation period of the star  (Vidotto et al, 2014). 
   It is commonly thought that the observed magnetic fields of young
stars arise from dynamo processes inside the rotating convective stars.
    However, it is possible that dynamo processes operate in both
 the accretion disc and in the star as considered by von Rekowski
 and Brandenburg (2006).  One then has the  possibility that the
 magnetic field generated in the disc is accreted onto the star and
 thereby modifies the stellar field changing the star's dipole, quadrupole,
 and octupole moments for example (Dyda et al. 2015). Time dependent dynamo processes could also be responsible for driving periodic or quasi-periodic outflows (Stepanovs, Fendt \& Sheikhnezami 2014).
 It is therefore important to understand the interaction 
 of the  dipole field of a rotating star with the disc  field. 
    In particular, we are interested in how the symmetry of the
stellar plus disc field configuration affects the symmetry or
asymmetry  of the outflow or jets.

   We consider four classes of magnetic field configurations - a pure disc field (DISC), a pure stellar dipole field (DIP), a disc field with anti-aligned stellar dipole field (ADIP) and a disc field with parallel aligned stellar dipole field (PDIP). 

   In DISC runs we consider an initial  reference {\it  disc-field}  the same as that studied   by Zanni et al. (2007) in axisymmetric FLASH code simulations which
assumed symmetry about the disc plane.
These simulations exhibited a stable, symmetric inner disc wind. 
    Later work by Fendt and Sheikhnezami (2013) considered the same
disc field  but without assuming symmetry about the equatorial plane. 
They found that this field gave rise to symmetric outflows,  but that
asymmetric outflows occurred if the disc density and temperature
were asymmetric about the equatorial plane.

    In addition to the mentioned disc-field, we consider accretion
onto magnetized rotating
stars with an aligned dipole magnetic field with magnetic moment $\mu$.
 Previous work by Lovelace et al. (2010) found that a pure dipole in the ``propeller regime'' exhibits asymmetric outflows. 
      In the propeller regime, the radius of the star's magnetosphere
 $R_{\rm m}$ (where the pressure of the disc matter is of the order of the magnetic energy density of the star's dipole field) is larger than the corotation 
 radius $R_{\rm cr}=(GM/\Omega_*^2)^{1/3}$ (where the
 rotation rate of the star equals the Keplerian rotation
 rate of the disc)  (Illarionov \& Sunyaev 1975).    In the
 propeller regime a large fraction of the accreting disc matter
 may be expelled from the system (Lovelace, Romanova \&
 Bisnovatyi-Kogan 1999; Romanova et al. 2004, 2005; Ustyugova
 et al. 2006; Lii et al. 2014).   A qualitative picture of
 the accretion is that matter accumulates at the magnetospheric boundary and compresses the magnetosphere inward until $R_m < R_{cr}$ and
funnel flow accretion occurs. 
   The funnel flow spontaneously breaks the system symmetry
about the equatorial plane and strong outflows are generated in the opposite hemisphere (Lii
et al. 2014). 
As the matter reservoir at the magnetospheric boundary is depleted, the 
magnetosphere expands outwards and the system can ``forget" which side had a funnel flow. 
     This process repeats on time scales of roughly 40 days for CTTS, with the side
determining the funnel flow being randomly determined, leading to episodic, asymmetric outflows (Lii et al. 2014). The DIP simulations are used to compare our simulations with stellar dipoles to these previous works.

Moreover, we consider the two cases where the stellar dipole field at the
magneotspheric radius is  parallel (PDIP) or anti-parallel (ADIP) to the disc-field.  
   This offers two types of behaviors which can  break the symmetry
about the equatorial plane.
   In the anti-parallel case (ADIP), the stellar dipole field may
reconnect  with the disc field as it is advected inward.   Analogous
reconnection occurs between the solar wind magnetic field
and the magnetic field of the earth when the two fields are
anti-parallel (e.g., Hargreaves 1992). This reconnection occurs at the magnetopause (inside the Earth's bow shock) where the subsonic solar wind runs inot the Earth's magnetosphere.
	In the parallel case (PDIP), the disc field causes additional pressure at the magnetospheric boundary, reducing the magnetospheric radius and altering the field line geometry outside the magnetosphere.

In  Sec. 2 we discuss the numerical methods used and the simulation parameters.  Section 3 discusses the four main cases simulated. Section 3.1 describes results  for the pure 
disc-field (DISC)  for different disc densities or plasma $\beta$ values. Section 3.2 describes results for a pure stellar dipole field (DIP), a disc-field and a parallel stellar dipole (PDIP) and a disc-field and an anti-parallel stellar dipole (ADIP). Section 4 gives the main conclusions from this work and its implications for observations.

\section{Theory}

\subsection{Basic Equations}

    The plasma flows are assumed to be described by the
equations of non-relativistic magnetohydrodynamics (MHD). 
In a non-rotating reference frame the equations are
\begin{subequations}
 \begin{equation}
\frac{\partial \rho}{\partial t} + \nabla \cdot \left( \rho \mathbf{v} \right) = 0~, 
\end{equation}
\begin{equation}
 \frac{\partial \rho \mathbf{v}}{\partial t} + \nabla \cdot \mathcal{T} = \rho \mathbf{g}~,
\end{equation}
\begin{equation}
\frac{\partial \mathbf{B}}{\partial t} + c\: \nabla \times \mathbf{E} = 0~,
\end{equation}
\begin{equation}
\frac{\partial \left( \rho S \right)}{\partial t} + \nabla \cdot \left( \rho  \mathbf{v}S \right) = \mathcal{Q}~.
\end{equation}
\end{subequations}
Here, $\rho$ is the mass density, $S$ is the specific entropy, $\mathbf{v}$ is the flow velocity, $\mathbf{B}$ is the magnetic field, $\mathcal{T}$ is the momentum 
flux density tensor, $\mathbf{E} = - \mathbf{v} \times \mathbf{B}/c + \eta_t \nabla \times \mathbf{B}/c$ is the electric field, $\mathcal{Q}$ is the rate of change of entropy per unit volume due to viscous
and Ohmic heating in the disc, $c$ is the speed of light and $\eta_t$ is the turbulent magnetic diffusivity. 
    We assume that the heating is offset by radiative cooling so that
$\mathcal{Q}=0$.
   Also, $\mathbf{g} = -\left(GM/r^2 \right)\hat{\bf r}$
is the gravitational acceleration due to the
central mass $M$.
We model the plasma as a non-relativistic ideal gas with equation of state
\begin{equation}
 S = \ln\left( \frac{p}{\rho^{\gamma}}\right)~,
\end{equation}
where $p$ is the pressure and $\gamma = 5/3$.

In most of this paper we use cylindrical coordinates $(R,\phi,Z)$ as these are the coordinates used by our code. 
However, for some purposes spherical coordinates are advantageous,
and they are denoted $(r,\theta,\phi)$.

   Both the viscosity and the magnetic diffusivity of the
disc plasma are thought to be due to turbulent fluctuations of the velocity and magnetic field.   
      Outside of the disc,  the
plasma is considered ideal with negligible viscosity and diffusivity.
      The turbulent coefficients are parameterized using
the  $\alpha$-model of Shakura and Sunyaev (1973).
    The turbulent kinematic viscosity is
\begin{equation}
 \nu_t = \alpha_{\nu} \frac{c_s^2}{\Omega_K}~,
\end{equation}
where $c_s$ is the midplane  sound speed, $\Omega_K$ is the Keplerian angular velocity at the given radii and $\alpha_{\nu}\leq 1$ is a dimensionless constant.
 Similarly,  the turbulent magnetic diffusivity is
\begin{equation}
 \eta_t = \alpha_\eta \frac{c_s^2}{\Omega_K}~,
\end{equation}
where $\alpha_\eta$ is another dimensionless constant. 
   The ratio, 
\begin{equation}
 \mathcal{P} = \frac{\alpha_{\nu}}{\alpha_{\eta}}~ ,
\end{equation}
is the magnetic Prandtl number of the turbulence in the disc
which is expected to be of order unity (Bisnovatyi-Kogan
\& Ruzmaikin 1976). 
   Shearing box simulations of MRI driven MHD turbulence
in discs indicate that ${\cal P}\sim 1$ (Guan \& Gammie 2009, Lesur \& Longaretti 2009) and this condition has been used in previous disc wind studies (see for example Tzeferacos et al. 2013). All our simulations use $\alpha_{\nu} = \alpha_{\eta} = 0.1$.

The momentum flux density tensor is given by
\begin{equation}
 \mathcal{T}_{ik} =p\delta_{ik}+ \rho v_i v_k  + \left( \frac{\mathbf{B}^2}{8\pi}\delta_{ik} - \frac{B_i B_k}{4\pi} \right) + \tau_{ik}~,
\end{equation}
where $\tau_{ik}$ is the viscous stress contribution from the turbulent fluctuations of the velocity and magnetic field. 
    As mentioned we assume that these can be represented in the same
way as the collisional viscosity by substitution of the turbulent viscosity.       Moreover,  under most conditions found in our simulations
$|v_Z| \ll |v_R| \ll v_{\phi}$  within the disc.
     Therefore we assume that the viscous stress is dominated mainly by terms involving the angular velocity gradient and the radial accretion velocity and choose to neglect terms involving $v_Z$.  
   The leading order contribution to the momentum flux density from 
turbulence is therefore

\begin{align}
\tau_{R\phi} = -\nu_t \rho R \frac{\partial \Omega}{\partial R} ~, \hspace{1.2cm}
\tau_{Z \phi} = - \nu_t \rho R \frac{\partial \Omega}{\partial z}~, \nonumber \\
 \tau_{Z R} = - \nu_t \rho \frac{\partial v_R}{\partial Z}~, \hspace{1.2cm}
\tau_{R R} = - 2 \nu_t \rho \frac{\partial v_R}{\partial R}~, \nonumber \\
 \tau_{\phi \phi} = - 2 \nu_t \rho \frac{v_R}{R} ~, \hspace{4cm}
\end{align}
where $\Omega = v_{\phi}/R$ is the plasma angular velocity.

    The transition from the viscous-diffusive disc to the ideal plasma
corona is handled  by multiplying the viscosity and
diffusivity by a dimensionless factor $\xi(\rho)$ which varies
smoothly from $\xi=1$ for $\rho \geq \rho_d=0.75\rho(R,Z=0)$   to
$\xi =0$ for $\rho\leq 0.25 \rho_d$ as described in Appendix B
of Lii, Romanova, \& Lovelace (2012).

   The MHD equations are written in dimensionless form 
so that the simulation results 
can be applied  to different types of stars. We assume that the central object is a CTTS with mass $M_* = 0.8 M_{\odot}$, a radius $R_* = 2R_{\odot}$ and a magnetic field magnitude $B_* = 3 \times 10^3$ G which is typical for the magnitude of the stellar dipole. We define dimensionful quantities with a 0 subscript, to denote typical values of physical parameters at a reference radius $R_0$.
    The mass scale is set by the stellar mass so  
$M_0 = M_*$. The reference length, $R_0 = 0.1 \rm{AU}$, is taken to
 be the scale of a typical inner disc radius.
Assuming a stellar dipole field, the magnetic field strength $B_0 = B_*/\mu (R_*/R_0)^3$. The mass, length and magnetic field scales allow us to define all other dimensionful quantities.

    The reference value for the
velocity is the Keplerian velocity at the radius $R_0$, 
$v_0 = (GM_0/R_0)^{1/2}$.  
     The
reference time-scale is the period of rotation at $R_0$, 
 $P_0 = 2\pi R_0/v_0$.
      From the MHD equations, we get the
relation $\rho_0 v_0^2 = B_0^2/4 \pi$,
where $\rho_0$ is a reference  density. This allows us to define the reference density $\rho_0$ of the disc.
    The reference mass accretion rate is $\dt{M}_0 = 4 \pi \rho_0 v_0 R_0^2$. The reference temperature $T_0 = v_0^2/2 \times m_H/k_B$ where $m_H$ is the atomic mass of hydrogen and $k_B$ is the Boltzmann constant is obtained by taking the ratio of the reference pressure and reference density. One can obtain dimensionful quantities from simulation results by multiplying by the appropriate dimensionful quantity in Table \ref{table:units}
    
\begin{table}
\begin{center}
  \begin{tabular}{ | l  c  l |}
                                                                 \\\hline
Parameters           & Symbol    & Value                      \\ \hline \hline
   mass              & $M_0$       & $1.59\times10^{33}$ g       \\
   length            & $R_0$       & $1.50\times10^{12}$cm        \\
   magnetic field    & $B_0$       & $8.04\times10^{-2}$ G                 \\ \hline 
   time              & $P_0$       & $3.53\times10^{-2}$ y     \\
   velocity          & $v_0$       & $8.42\times10^{7}$cm/s    \\ 
   density           & $\rho_0$    & $7.24\times 10^{-15} $ g/cm$^3$\\
   accretion rate    & $\dt{M}_0$  & $2.72\times 10^{-7}$ $M_\odot$/yr \\
   temperature       & $T_0$       & $4.26\times 10^{7}$ K \\
   dipole strength   &$\mu$        & $2.69\times 10^{35}$ $\rm{G \ cm}^3$\\
        \hline \hline
  \end{tabular}
\end{center}
\caption{Mass, length, and magnetic field scales of interest and the corresponding scales of other derived quantities for a CTTS. One can obtain the physical values from the simulation values by multiplying by the corresponding dimensional quantity above.}
\label{table:units} 
\end{table}
Results obtained in dimensionless form can be applied to objects with 
widely different sizes and masses. 
    However, the present work focuses on CTTS with the
typical values shown in Table \ref{table:units}. A more detailed description of our code and numerical methods can be found in Koldoba et al. (2015).

\subsection{Initial Conditions}

\subsubsection{Magnetic Field}

   The initial magnetic fields are taken to be  linear combinations 
of the disc-field (Zanni et al. 2007) defined by the flux function
\begin{equation}
 \Psi_{d} = \frac{4}{3} B_0 R_0^2 \left( \frac{R}{R_0} \right)^{3/4} \frac{m^{5/4}}{\left( m^2 + Z^2/R^2\right)^{5/8}}~,
\end{equation}
and a  fixed dipole field centered in the star which
has the flux function
\begin{equation}
 \Psi_D = \frac{\mu R^2}{\left(R^2 + Z^2 \right)^{3/2}}~,
\end{equation}
where $\mu=$ const is the star's dipole moment. 
Our base case is taken to be $\mu = 30$,
but  we also consider cases with $\mu = 10, 20$.
    The  dimensionless parameter $m$ in Eqn. 8  determines 
the bending of the disc-field lines in the poloidal plane with the limit $m \rightarrow \infty$ corresponding to purely vertical field lines. We assume $m = 1$.
     The poloidal magnetic field components are given by
\begin{equation}
 B_Z = \frac{1}{R} \frac{\partial \Psi}{\partial R}, \hspace{2cm} B_R = - \frac{1}{R} \frac{\partial \Psi}{\partial Z}~.
\end{equation}
       The relative strengths of the disc and dipole
  field components are characterized by the
radius $R_B$ where the stellar dipole 
and disc-field have equal values in the midplane of the disc at $t = 0$. 
    We take $R_B = 2.8, 4.1, \&~ 5$.   
    The magnetospheric radius $R_m$ is determined by plotting the angular velocity
of the matter in the equatorial plane as a function of $R$.  The change
from the constant stellar rotation to Keplerian occurs at $R_m$.
       As mentioned in \S 1, the propeller regime occurs for $R_{m} > R_{\rm cr}$.

\begin{figure*}
                \centering
                \includegraphics[width=\textwidth]{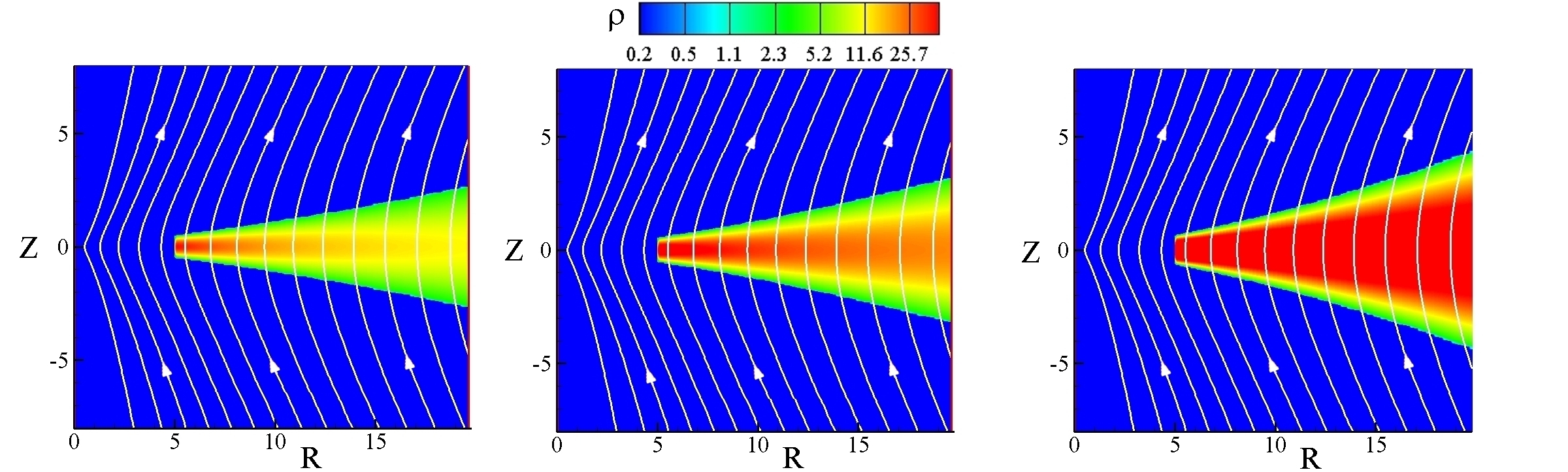}
        \caption{The initial density $\rho$ and poloidal field lines for radially
  distributed disc-field in the cases of Bernoulli parameters \emph{Left:} ${\cal B}_0 = 3 \times 10^{-4}$  \emph{Center:} ${\cal B}_0 = 5 \times 10^{-4}$, and 
\emph{Right:} ${\cal B}_0 = 1 \times 10^{-3}$. 
   Increasing the Bernoulli parameter increases the midplane disc density and the disc scale height. }
\label{fig:zanni_ic}
\end{figure*}

\begin{figure*}
                \centering
                \includegraphics[width=\textwidth]{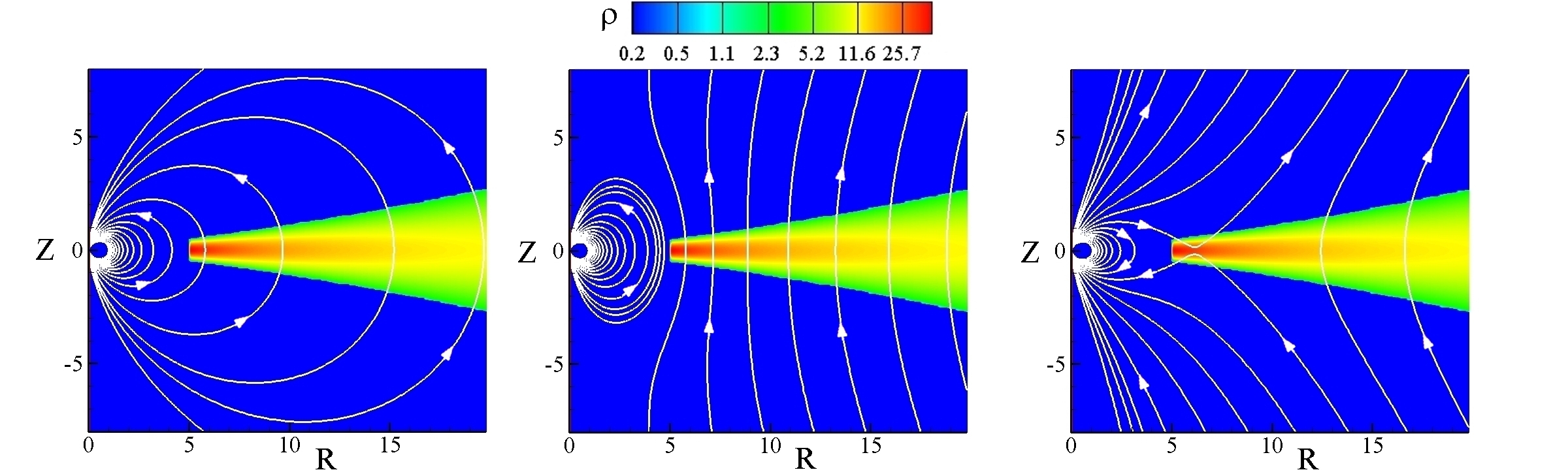}
        \caption{Plot of density $\rho$ and poloidal field lines for a 
 stellar dipole (left), a stellar dipole and parallel disc-field (center), and a stellar
 dipole and an anti-parallel disc-field.}
\label{fig:dipole_ic}
\end{figure*}

Our simulations cover a wide range of dipole and disc-field
combinations. In particular, we consider a pure  disc-field (DISC), a pure stellar dipole (DIP), a dipole and parallel disc-field (PDIP), and a dipole with an anti-parallel disc-field (ADIP). 
  Figure \ref{fig:zanni_ic} shows the initial disc density and
poloidal field lines for the disc-field DISC. 
Figure \ref{fig:dipole_ic} shows the same quantities for cases
including an aligned stellar dipole magnetic field (DIP, PDIP \& ADIP).

\subsubsection{Matter Distribution}

Initially, matter of the disc and corona are assumed to be in mechanical equilibrium (Romanova et al. 2002). 
The initial density distribution is taken to be barotropic with
\begin{equation}
  \rho(p) =
  \begin{cases}
   p/T_{\rm{disc}}, & p>p_b ~~{ \rm and} ~~ R \geq R_d~, \\
   p/T_{\rm{cor}}, & p<p_b ~~ {\rm or} ~~ R  \leq R_d~,
  \end{cases}
\end{equation}
where $p_b$ is the level surface of pressure that separates the cold matter of the disc from the hot matter of the corona and $R_d$ is the inital inner disc radius. The initial inner radius of the disc is $R_d = 5 R_0$.   At this surface
the density has an initial step discontinuity from value $p/T_{\rm{disc}}$ to $p/T_{\rm{cor}}$. The initial dimensionless temperature in the disc is $T_{\rm{disc}} = (p/\rho)_{\rm{disc}} = 5\times10^{-4}$, and the initial temperature in the corona is $T_{\rm{cor}} = 
(p/\rho)_{\rm{cor}} = 0.5$.

Because the density distribution is barotropic, the initial angular velocity is a constant on coaxial cylindrical surfaces about the $z-$axis. Consequently, the pressure 
can  be determined from the Bernoulli  equation
\begin{equation}
 F(p) + \Phi + \Phi_c = {\cal B}_0~,
\label{eq:forcebalance}
\end{equation}
where ${\cal B}_0$ is Bernoulli's constant,$\Phi = -GM/\sqrt{R^2 + Z^2}$ is the gravitational potential with $GM = 1$ in the code, $\Phi_c = \int_{R}^{\infty}r dr ~\Omega^2(r)$ is the 
centrifugal potential, and
\begin{equation}
  F(p) =
  \begin{cases}
   T_{\rm{disc}}\ln(p/p_b) & p>p_b ~~ {\rm and} ~~ R \geq R_d~, \\
   T_{\rm{cor}}\ln(p/p_b) & p<p_b ~~{\rm or} ~~ R \leq R_d~.
  \end{cases}
\end{equation}
    Varying ${\cal B}_0$ allows one  to vary the initial midplane disc density. We consider three values of the Bernoulli parameter ${\cal B}_0 = 3 \times 10^{-4}, 5 \times 10^{-4} \ \rm{and} \ 10^{-3}$, which we refer to as the diffuse, intermediate and dense disc cases respectively. 
        The initial half-thickness of the disc $h$ is taken to be  $h/R = 0.100$ at the inner disc radius $R_d = 5$. 

\begin{figure}
                \centering
                \includegraphics[width=.5\textwidth]{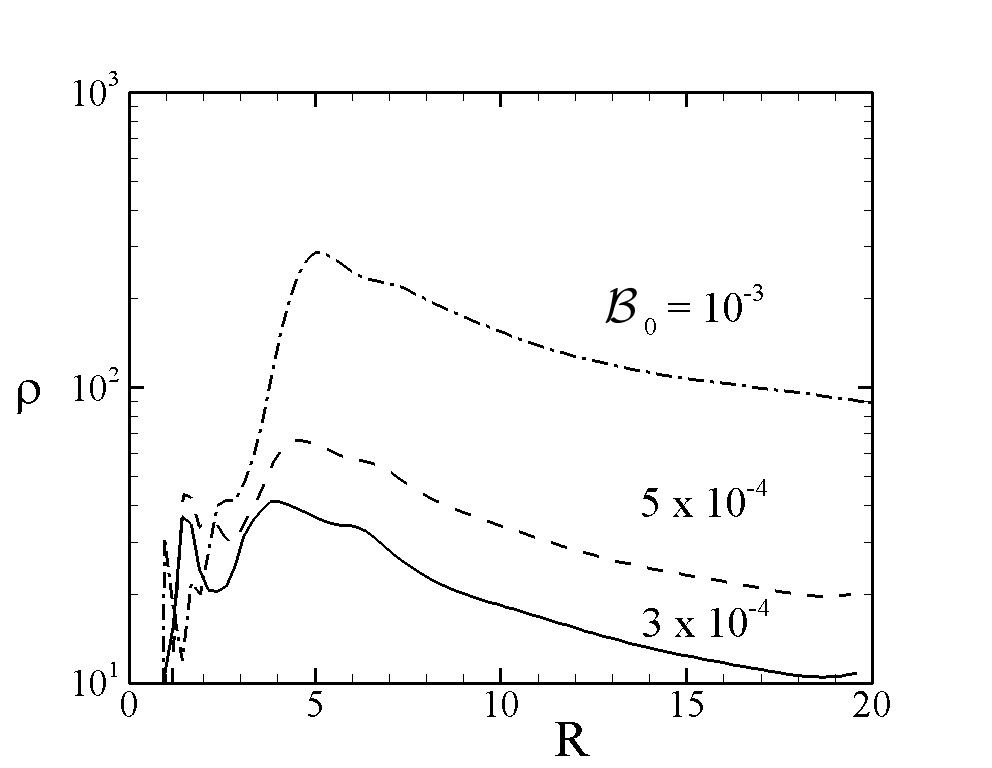}
        \caption{Initial midplane density profile $\rho(R,0)$ as a function of radius for different values of the Bernoulli parameter
${\cal B}_0 = 3 \times 10^{-4}, 5 \times 10^{-4} \ \rm{and} \ 10^{-3}$.} 
\label{fig:zanni_density}
\end{figure}

\begin{figure}
    \centering
    \includegraphics[width=0.5\textwidth]{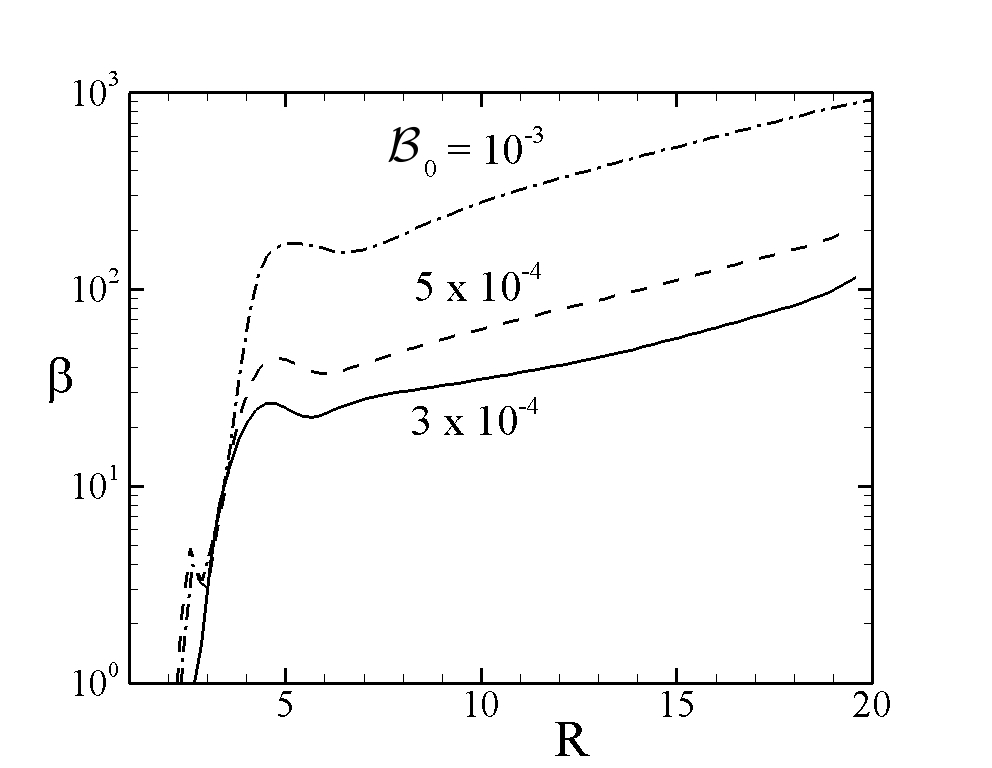}
 \caption{Initial radial dependences of the midplane $\beta =  8\pi p/B^2$ for
 the different Bernoulli parameters.}
 \label{fig:zanni_beta}
\end{figure}

      Figure \ref{fig:zanni_density} shows
 the radial dependence of the midplane disc density for
 the different Bernoulli parameters.   
Figure \ref{fig:zanni_beta} shows the initial radial dependence of
the midplane $\beta =  8\pi p/B^2$ (the ratio of the gas
pressure to the magnetic pressure) for the different Bernoulli parameters.

\begin{figure}
                \centering
                \includegraphics[width=0.5\textwidth]{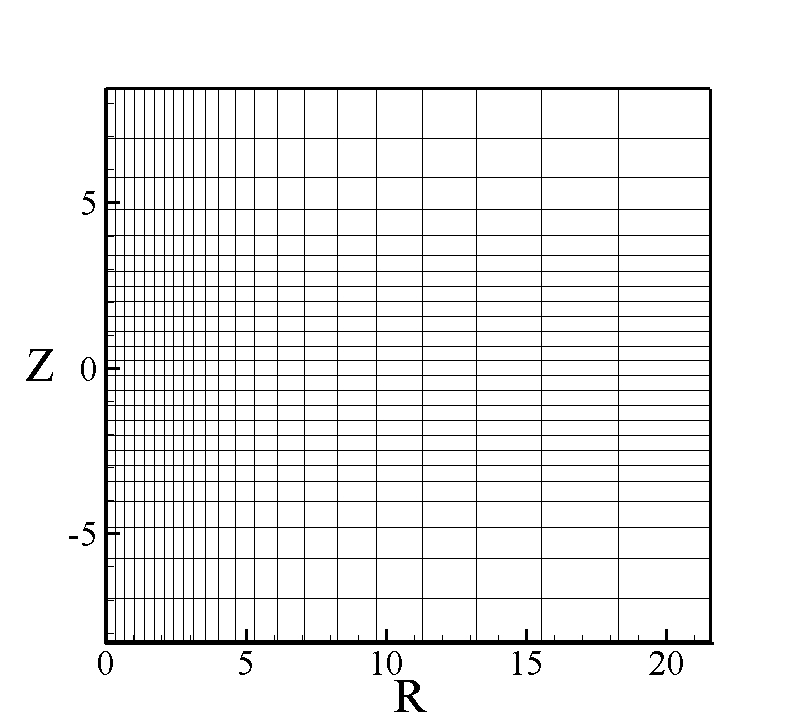}
        \caption{Sparse version of grid used in the simulations, showing every $7$th gridline in the $R-$direction and every $9$th gridline in the $Z-$direction.}
\label{fig:mesh}
\end{figure}

\subsubsection{Angular Velocity}

     The initial angular velocity of the disc is slightly sub-Keplerian, 
\begin{equation}
 \Omega = (1-0.003)\Omega_K(R) \hspace{1cm} R>R_d~,
\end{equation}
Inside of $R_d$, the matter rotates rigidly with angular velocity
\begin{equation}
 \Omega = (1-0.003)\Omega_K(R_d) \hspace{1cm} R  \leq R_d.
\end{equation}
The corotation radius $R_{\rm cr}$ is the radius where
the angular velocity of the disc equals that of the star; that is,
$R_{\rm cr}=(GM/\Omega_*^2)^{1/3}$.  
In this study we have chosen $R_{\rm cr} =1.5$. 
     For the considered stellar dipole fields and accretion rates, the
systems are in the strong propeller regime where $R_m$ is significantly
larger than $R_{\rm cr}$.

\subsection{Boundary Conditions}

Our simulation region has three boundaries: the axis, the surface of the star and the external boundaries. For each dynamical variable we impose a boundary
condition consistent with our physical assumptions.

We assume axisymmetry about the axis. On the star and the external boundaries we want to allow fluxes and impose free boundary conditions
$\partial \mathcal{F}/\partial n = 0$ where $\mathcal{F}$ is a dynamical variable and $n$ is the vector normal to the boundary. 

At the external boundary along the edge of the disc we allow new matter to flow into the simulation region. We impose the condition that the matter
must be accreting $v_r < 0$. In the coronal region, we prescribe outflow conditions and allow matter, entropy and magnetic flux to exit the simulation region.  

The star is taken to be cylindrical in shape with radius $R_* = 1$ and height $Z_* = 2$. Free boundary conditions are imposed on the variables $\rho$, $p$ and $\mathbf{B}$. The velocity vectors are required to align with the magnetic field lines on the star via $\mathbf{v} = (\mathbf{B}\cdot \mathbf{v})\mathbf{B}/|\mathbf{B}|$. In addition, we require that matter on the stellar boundary flow into the star. We treat the corner of the star by averaging over the nearest neighbour cells in the R and Z directions.

In addition to the boundary conditions imposed for the well-posedness of our problem we impose additional conditions on variables at the boundaries to eliminate numerical
artifacts of the simulations. 

Our simulation uses a grid of $154 \times 225$ cells. The star has a radius of 1 in units of the simulation and is cylindrical in shape. It extends
10 grid cells above and below the equatorial plane. In the R-direction, 
the first 30 grid cells have length $dR = 0.05$. Later cell lengths are given recursively by $dR_{i+1} = 1.025 dR_{i}$. Similarly, in the Z-direction
the first 30 grid cells above and below the equatorial plane have length $dZ = 0.05$. Later cell lengths are given recursively by $dZ_{j+1} = 1.025 dZ_{j}$.   Figure \ref{fig:mesh} shows
a sparse version of our grid with only every $7$th gridline in the 
$R-$direction and every $9$th gridline in the $Z-$direction are shown.

\section{Results}
We are interested in the mass, angular momentum, and energy  outflow
from the disc.
      We have also calculated these fluxes at larger values
  of $|Z|$ and found this did not significantly change the values.
     We argue it is best to avoid the last few simulation region cells to avoid boundary effects while being as far away from the star as possible to allow the matter to be accelerated by the magnetic field. 
     We note that mass fluxes are more symmetric when measured near the surface of the disc but this is observationally less interesting than the flux at large Z. 
      In addition, for the mass flux we require that $v_z > 0.1 $ which corresponds to an observationally interesting value of $\gtrsim 200$km/s. A summary of our runs, their key parameters and results are found in Table \ref{table:summary}.

\begin{table}
\begin{center}
  \begin{tabular}{ l  l  l  l  l}
                                                                 \\ \hline
Type & $\mu$ & $B_0$ & $\mathcal{B} (10^{-4})$ & Outflow Summary \\ \hline \hline
     &   0   &   1  &      3                   & Long lasting, symmetric \\
DISC &   0   &   1  &      5                   & Quasi-periodic\\
     &   0   &   1  &     10                   & Persistently asymmetric \\
\hline
     &  10   &   0  &      3                   & Strong disc/mag. wind\\
DIP  &  20   &   0  &      3                   & Strong disc/mag. wind\\
     &  30   &   0  &      3                   & Strong disc/mag. wind \\
\hline
     &  10   &   1  &      3                   & No magnetospheric wind \\
PDIP &  20   &   1  &      3                   & Weak disc/strong mag. wind \\
     &  30   &   1  &      3                   & Weak disc/strong mag. wind \\
\hline
     &  10   &  -1  &      3                   & Suppressed \\
ADIP &  20   &  -1  &      3                   & Suppressed \\
     &  30   &  -1  &      3                   & Suppressed \\
\hline \hline
  \end{tabular}
\end{center}
\caption{Summary of key parameters in our runs and a qualitative summary of the outflows observed.}
\label{table:summary} 
\end{table}

\subsection{Pure Disc-Field DISC}

In this series of runs we varied the Bernoulli parameter ${\cal B}_0$ which determines the initial matter configuration (Eqn. 12). 
   We take  ${\cal B}_0 = 3 \times 10^{-4}, 5 \times 10^{-4} \ \rm{and} \ 10^{-3}$. 
       Increasing the Bernoulli parameter increases the midplane density, and therefore the total mass of the disc. 
       The initial midplane density profiles are shown in Fig. \ref{fig:zanni_density}.  For simplicity, we  refer to them as the diffuse or low-$\beta$  (${\cal B}_0 = 3 \times 10^{-4}$), intermediate (${\cal B}_0 = 5 \times 10^{-4}$), and dense or
high-$\beta$ (${\cal B}_0 = 10^{-3}$) discs, respectively.

    For the same disc-fields, the three disc configurations exhibit qualitatively different behaviors of their outflows. The diffuse low-$\beta$ disc exhibits symmetric ($\pm Z$), long lasting outflows as observed by Fendt and Sheikhnezami (2013). 
     The intermediate disc exhibits oscillatory behaviour, with the  jet flip-flopping
several times before settling at late time to steady symmetric outflows. 
     The dense  high-$\beta$ disc exhibits outflows which are  persistently asymmetric.  This is the result of the excess accumulation of
matter in one or the other of the two hemispheres near the magnetosphere which
then feeds more matter into one or the other outflows. Murphy, Ferreira and Zanni (2010) have shown that the important criteria for launching outflows is that $\beta \sim 1$ at the disc surface. Our results show that the character of these outflows is affected by the midplane $\beta$.

\subsubsection{Diffuse Low-$\beta$ Disc}

\begin{figure}
                \centering
                \includegraphics[width=.5\textwidth]{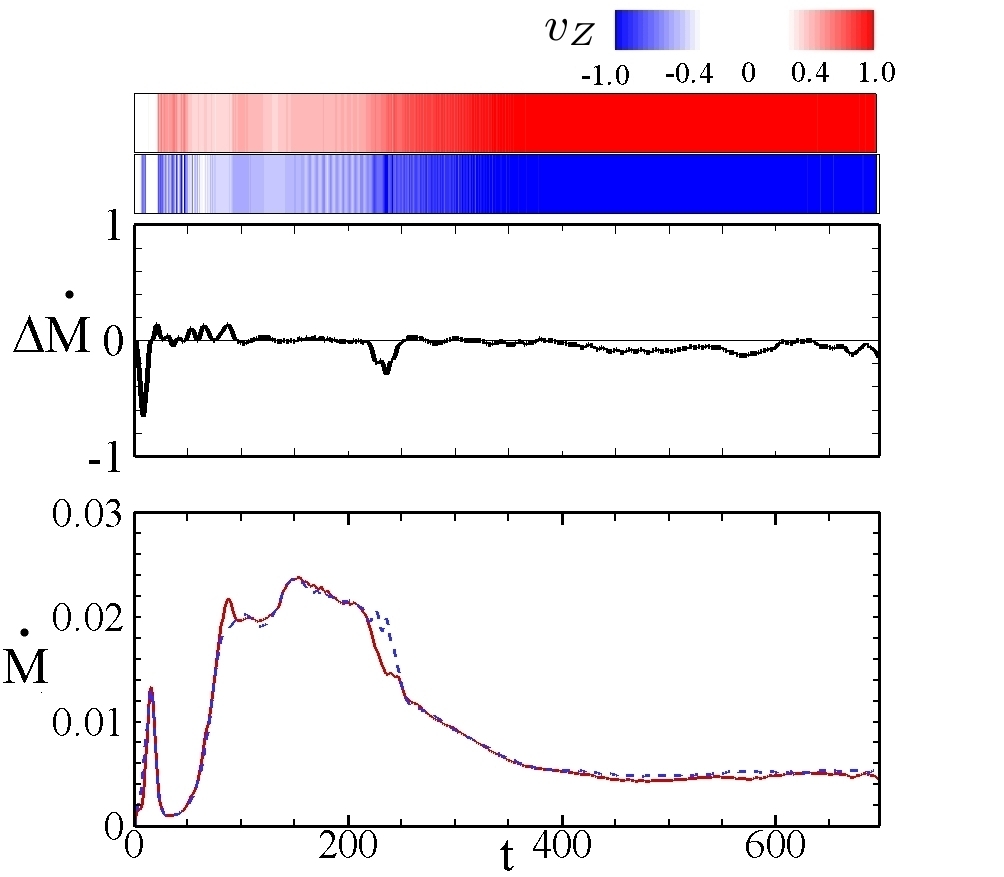}
        \caption{ {\it Diffuse  low$-\beta$ disc:} The lower plot shows the mass flux $\dot{M}$ through the upper (solid red line) and lower (dashed blue line) boundaries. 
        The middle plot shows the mass flux difference, normalized to average total mass flux $\Delta \dot{M} = (\dot{M}_t - \dot{M}_b)/(\dot{M}_t + \dot{M}_b) $ as a function of time. The upper plot shows the average velocity of outflows, weighted by mass along the upper (red) and lower (blue) boundaries. Both the mass flux and velocity of outflows is symmetric in this case.} 
\label{fig:mdot_thin}
\end{figure}

Figure \ref{fig:mdot_thin} shows
the mass flux $\dt{M}$ through the upper hemisphere (solid red
line) and lower hemisphere (dashed blue line) of the low $\beta$ disc as a function of time.
 This configuration exhibits outflows with mass fluxes of order 0.1\% of the stellar accretion rate. For ease in comparing the outflows, we plot the average mass flux asymmetry $\Delta \dt{M} = (\dt{M}_t - \dt{M}_b)/(\dt{M}_t + \dt{M}_b) $ as a function of time. We notice that $t < 400$ is characterized by sporadic asymmetries of order 10\% of the outflow rate. For $t > 400$ the asymmetry is still roughly 10\% but is persistently in the lower half plane. The upper contour plot shows the average velocity, weighted by mass, of outflows along the upper (red) and lower (blue) boundaries.

This serves as a calibration run in the respect that it reproduces the results of Fendt and Sheikhnezami 
 (2013) which found persistent, symmetric outflows for $t < 500$ in our simulation units. 
     Their outflow was measured near the surface of the disc, at a distance of $z = 3H$. We note that our outflows are more symmetric when measured at smaller heights, notably the small asymmetry near $t \sim 100$ vanishes. However, as we are ultimately interested in outflows which exceed the local escape velocity, we choose to measure outflows closer to the outer boundary to allow the outflows to be magnetically
  accelerated. We also note that both outflows are launched with comparable velocities.
 
\subsubsection{Intermediate Disc}

\begin{figure}
                \centering
                \includegraphics[width=.5\textwidth]{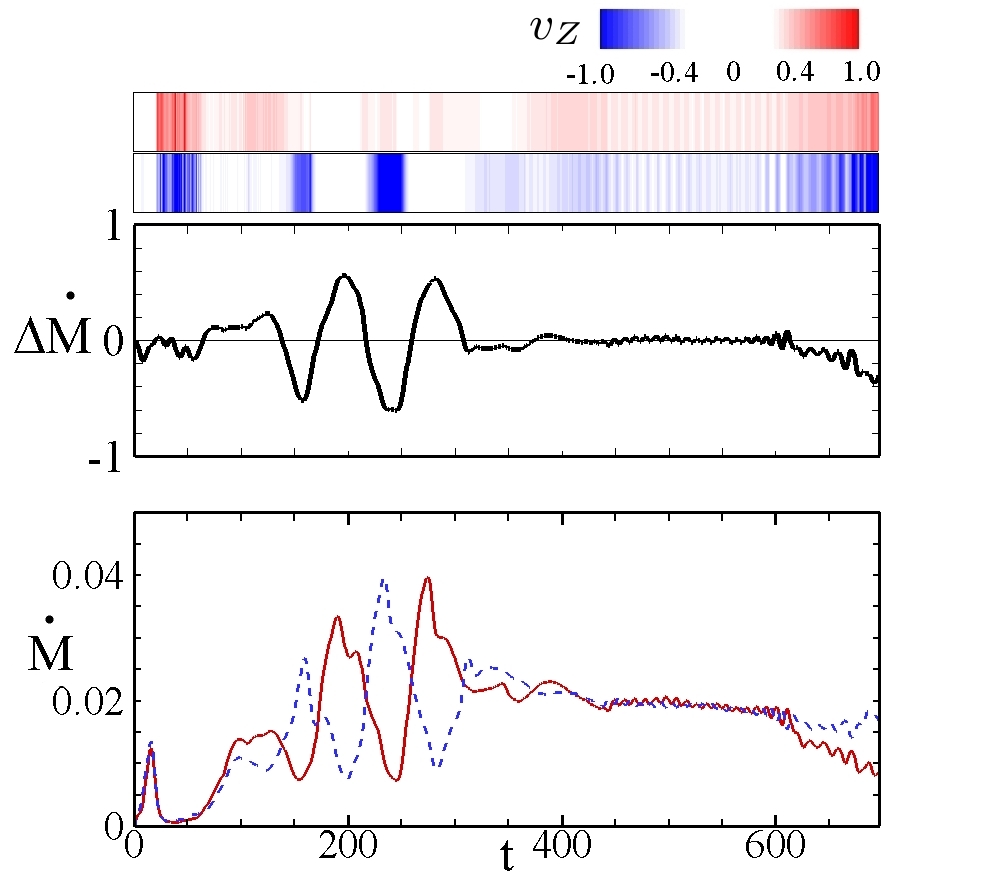}
        \caption{{\it Intermediate density disc:} The bottom panel shows the mass flux $\dot{M}$ through the upper (solid red line) and lower (dashed blue line) boundaries. The middle panel shows the mass flux difference, normalized to average total mass flux $\Delta \dot{M} = (\dot{M}_t - \dot{M}_b)/(\dot{M}_t + \dot{M}_b) $ as a function of time. The upper plot shows the average velocity of outflows, weighted by mass along the upper (red) and lower (blue) boundaries. Though the mass flux is symmetric during the oscillatory period $100 < t < 400$ the velocity of outflows is greater across the lower boundary.} 
\label{fig:mdot_intermediate}
\end{figure}

   Figure \ref{fig:mdot_intermediate} shows
the mass flux $\dt{M}$ through the upper hemisphere (solid red
line) and lower hemisphere (dashed blue line) as a function of time. 
 After an initial period of symmetric outflows, we observe for times $100 < t < 400$ quasi-periodic, oscillating, outflows. For times $t > 400$ the outflows are again symmetric with asymmetries on the order of a few percent.

\begin{figure*}
                \centering
                \includegraphics[width=\textwidth]{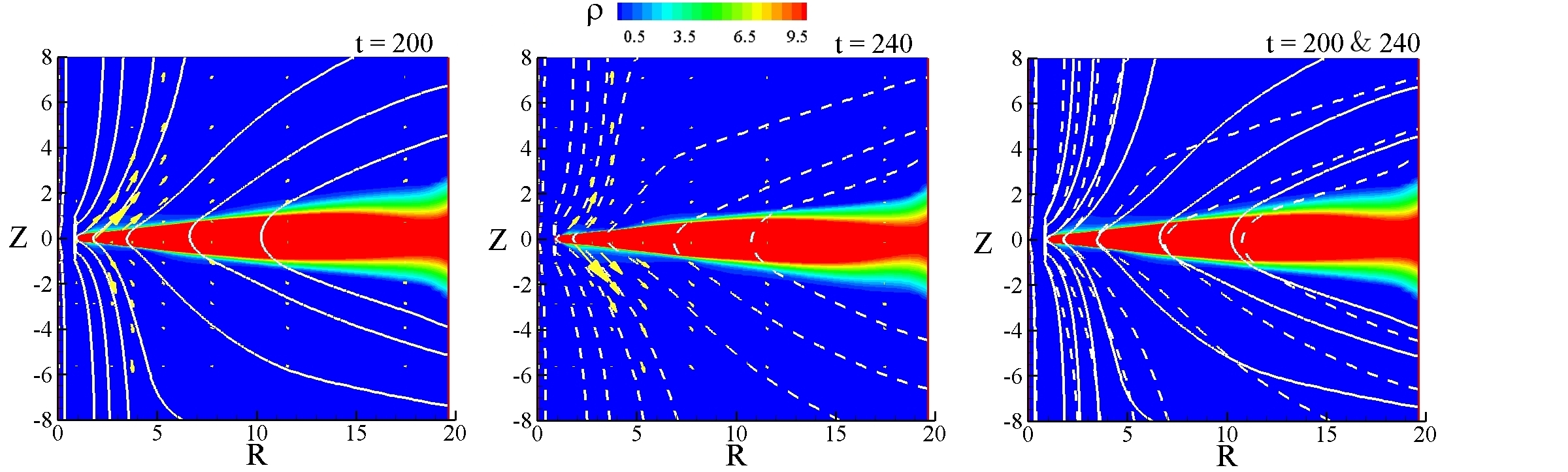}
        \caption{{\it Intermediate density disc:} Density $\rho$ (color), poloidal field lines $\Psi$ (white) and coronal mass flux $\rho {\bf v}_p$ (yellow).
      The  left-hand panel is at  $t= 200$, and the center panel is
at  $t = 240$. 
   The right-hand panel shows a superposition of the field lines
 at these two times which  corresponds to a maxima in the mass outflow in the upper  and lower hemispheres, respectively.
  Note that the field line structure exchanges between the two hemispheres at the two times, with the more inclined field lines being associated with stronger outflows.} 
\label{fig:oscillations}
\end{figure*}

   The oscillations in the outflows are caused by a periodic motion of the field lines. 
    In one hemisphere the field lines become more vertical, reducing the outflow, whereas in the other they become more inclined relative
 to the $Z-$axis.
      This increases the outflow, as matter is more readilly launched along inclined field lines (e.g., Lii et al. 2012).    As shown
in Fig. \ref{fig:oscillations} the structure then reverses, the more vertical field lines becoming more horizontal and vice versa.
      This leads to quasi-periodic oscillations in the mass outflows with a period  $T \sim 80$ and an amplitude of roughly $50\%$. 
                At the peak of the oscillation, there is therefore a factor of $\sim 3$ between the upper and lower mass flux. 
                For a T Tauri star with our choice of reference units this corresponds to variations on a time scale of roughly $2.8$ years. 
                The oscillations are caused by   Alfv\'en waves which
 propagate with a speed $v_A  \sim  \beta^{-1/2} c_s \sim \beta^{-1/2} h/R $.   The 
time-scale $\tau$ for passage of the wave across the disc a distance $\sim 2 h$
is $\tau \sim 2 \sqrt{\beta} R/v_K \sim \sqrt{\beta} R^{3/2} $. 
We find a midplane $\beta \approx 40$ and take $R \approx 3$, the radius from which strong outflows are launched and find $\tau \sim  60$, which agrees 
approximately with the observed period $T \sim 80$.

We note that though the peak mass flux through each hemisphere is comparable during this oscillatory phase, the velocity of the outflows is not. In particular, the bottom jets velocity $v_Z \approx - 0.1$ during mass flux minima but rises to $v_Z \approx -0.8$ and $-1.5$ during its two peaks. The top jet on the otherhand has $v_Z \approx 0.2$ during outflow minima and peaks at $v_Z = 0.3$ and $0.4$. A mass flux maxima of the bottom jet therefore could explain RW Aur A observations for instance, where the jet has a mass flux of 2-3 times the counter jet and a velocity roughly twice as great. 

At late times, $t > 600$, a persistent asymmetry of the outflows is observed. 
    This occurs because the original matter in the main part of the disc has been depleted and new matter is being added along the right boundary. 
    This induces an asymmetry in the outer part of the disc which in turn
induce disc oscillations that lead to a tilted disc.  This asymmetry appears
to be due to our outer boundary condition.

\subsubsection{Dense High-$\beta$ Disc}

A more massive disc results in a higher density of matter close to the star. At late times the disc half thickness h/R is roughly 5 times greater than for the diffuse disc in the very inner disc $R \lesssim 5$. The matter available for outflows is hence greater but is also more strongly influenced by inner disc oscillations. Radially propagating disc waves matter preferentialy accumulates in the upper or lower hemisphere leading to stronger outflows. However, unlike in the dipole case where the magnetosphere re-expands after a strong outflow, reestablishing a symmetric disc, these asymmetries persist until strong vertical disc oscillations develop or the asymmetric matter reservoir is depleted.

\begin{figure}
                \centering
                \includegraphics[width=.5\textwidth]{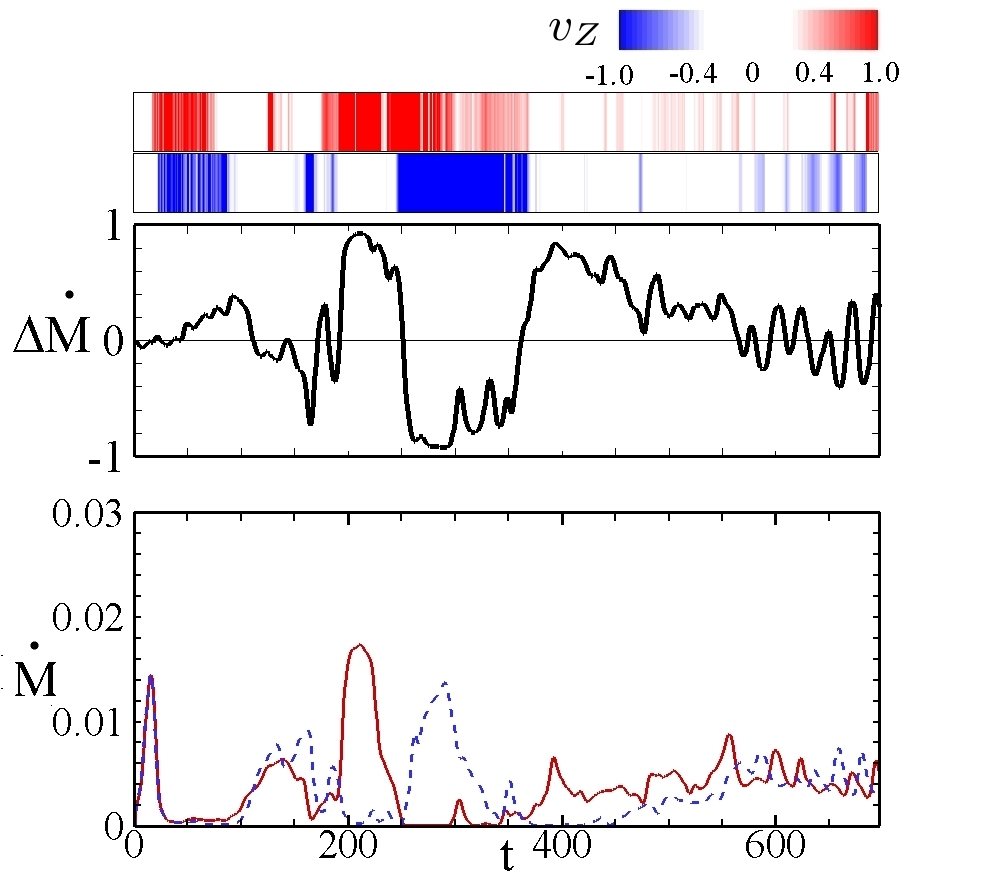}
        \caption{{\it Dense high-$\beta$ disc:} The bottom panel shows the mass flux $\dot{M}$ through the upper 
  (solid red line) and lower (dashed blue line) boundaries.  
         The middle panel shows the  mass flux difference, normalized to average total mass flux $\Delta \dot{M} = (\dot{M}_t - \dot{M}_b)/(\dot{M}_t + \dot{M}_b) $ as a function of time. The upper plot shows the average velocity of outflows, weighted by mass along the upper (red) and lower (blue) boundaries.} 
\label{fig:mdot_thick}
\end{figure}

As in the case of the intermediate disc,  there are strong oscillations of the inner disc field.
      These lead to spontaneous one sided outflows as the field lines become more 
 inclined to the $Z-$axis (Fig. \ref{fig:mdot_thick}). The outflows turn off as the field becomes more
 vertical. 
      However, unlike in the intermediate disc case these oscillations are not coupled
between the two hemispheres.
    The field lines in the top and bottom hemispheres  oscillate and rearrange themselves largely independently of one another. 
          For denser discs, poloidal field lines going through the disc are approximately frozen-into the disc which moves relatively slowly.  Consequently,  the field  motion above and below the disc are uncoupled, allowing the field lines to have greater curvature around the midplane of the disc. This is consistent with an observation made by Tzeferacos et al (2009), where simulations with discs with larger plasma $\beta$ supported field lines with greater curvature. These early time asymmetries last on time scales of order $T \sim 100 \sim 3.5 y$. In constrast, at late times asymmetries last on order of $T \sim 10 \sim 0.35 y$. This suggests that as the disc has a chance to reach a more stable configuration, timescales of asymmetries will decrease.

\subsection{Stellar Dipole}

  A rotating star with a dipole field in the propeller regime exhibits quasi-episodic, one sided outflows (Lovelace et al. 2010). 
  Matter builds up at the magnetospheric boundary, compressing the dipole until there is a symmetry breaking and the built up matter accretes 
to  the star in a one-sided funnel flow. 
The funnel flow leads to a one-sided outflow in the opposite hemisphere, after which the matter pressure is decreased, the magnetosphere re-expands and the process can begin anew.

To minimize the asymmetries induced by disc oscillations, this set of simulations uses the parameters for a diffuse disc. However,  the simulations with an intermediate disc 
showed no qualitative differences in the results. 
    We consider the simple case of a stellar dipole aligned with the rotation axis of the star. 
        In addition, there may be a disc-field.
        There are two possible configurations - one where the two fields are  parallel 
and the other where they are anti-parallel.

 In the anti-parallel configuration there is an X-type magnetic neutral line in the disc (see Fig. 2). As the large scale disc-field advects inwards, it annihilates with the dipole, allowing matter to accrete onto the star in a continuous fashion, different from the episodic matter inflows of a pure dipole. Despite the additional energy available to launch outflows due to the increased stellar accretion rate, the field line geometry does not result in enhanced outflows.

In a parallel configuration, the large scale field compresses the dipole, reducing the magnetospheric radius. The key parameters for determining the strengths of magnetospheric outflows is the relative position of the magnetospheric radius $R_m$ and the co-rotation radius $R_{\rm cr}$. When $R_{\rm cr} < R_m$ we are in the propeller regime and have strong magnetospheric outflows. 
     Keeping other parameters fixed, one can therefore imagine increasing the strength of the large disc-field and turning off the propeller. 
      A second type of outflow is an inner disc wind, magneto-centrifugally launched from open field lines threading the disc. Though the disc wind is launched over the entire disc, mass flux is dominated by outflows from the inner part of the disc. Therefore, increasing $R_m$ until it is at the radius of the inner disc wind significantly reduces this outflow, as the open field lines are pushed outwards.

For our base case we consider a stellar dipole with dimensionless 
magnetic moment  $\mu = 30$. With an anti-parallel disc-field of the type considered in the previous section, this configuration initially has an X type neutral line at $R = 6$ (see Fig. \ref{fig:dipole_ic}). 
   The neutral line should be at a radius large enough so that the dipole is not fully compressed by the disc and its effect essentially negated. However, since most outflows from the disc-field are generated at small radii, the neutral line should not be too far out either. 

In the parallel field configuration we consider dipoles with $\mu = 10, ~20,~ 30$. These correspond to initial magnetospheric radii of 
$R_m = 2.8, ~4.1, ~5$ respectively. The co-rotation radius
is $R_{\rm cr} = 1.5$ so that initially the system is in the strong propeller regime.  

\subsubsection{Pure Stellar Dipole Field DIP}

\begin{figure}
                \centering
                \includegraphics[width=.5\textwidth]{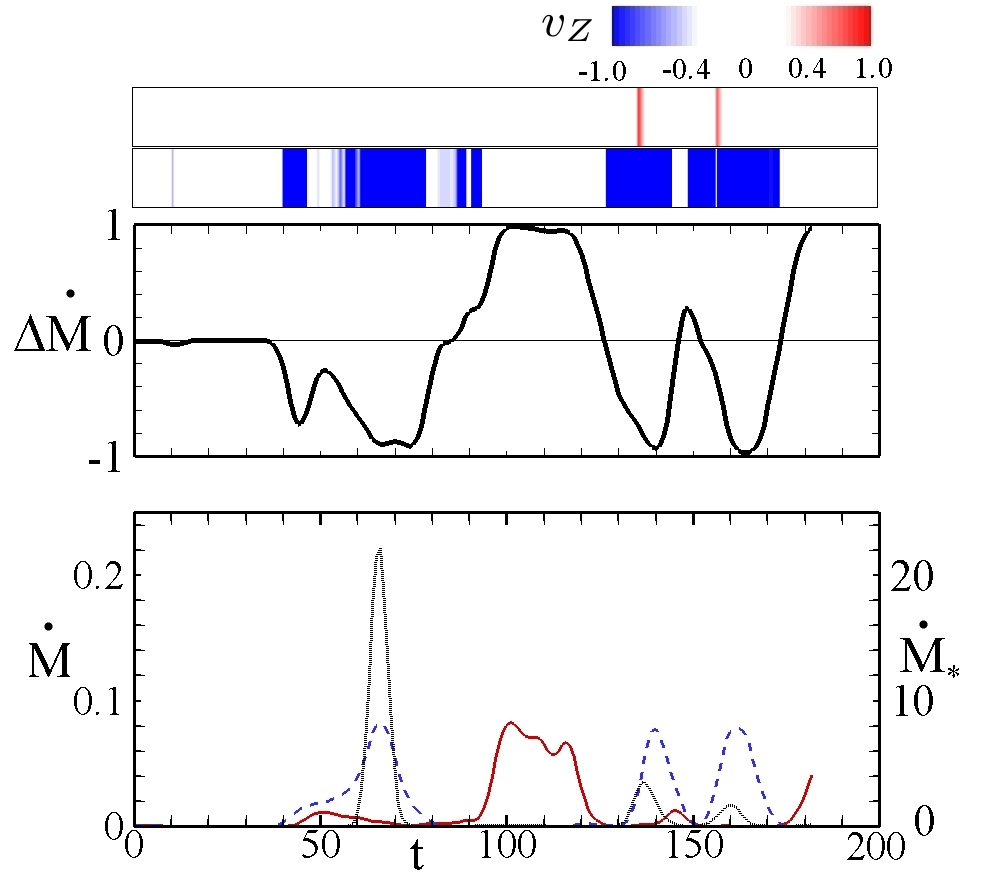}
        \caption{{\it Pure Stellar Dipole:} Lower panel shows stellar mass accretion rate $\dot{M}_*$ (dotted black line line) and mass outflow $\dot{M}$ (red solid line for upper hemisphere; blue dashed line
 for lower hemisphere) as a function of time. The middle panel shows the mass flux difference, normalized to average total mass flux $\Delta \dot{M} = (\dot{M}_t - \dot{M}_b)/(\dot{M}_t + \dot{M}_b) $ as a function of time. The upper plot shows the average velocity of outflows, weighted by mass along the upper (red) and lower (blue) boundaries.
       The right-hand vertical axis for the outflows has been scaled so $\dot{M}/\dot{M}_* = 10^{-2}$. Note the two types of outflows. 
        When the stellar accretion rate correlates to the matter outflow this corresponds to funnel flow accretion in one hemisphere and outflow in the other. The three such events correspond to funnel flow accretion in the upper hemisphere and ejections in the lower hemisphere. These outflows occur at fast velocity.} 
       Outflow when there is no stellar accretion corresponds to matter pushing against the magnetosphere in one hemisphere and being ejected outwards without 
       matter accretion to the star. 
\label{fig:pure_star+out}
\end{figure}

\begin{figure}
                \centering
                \includegraphics[width=.5\textwidth]{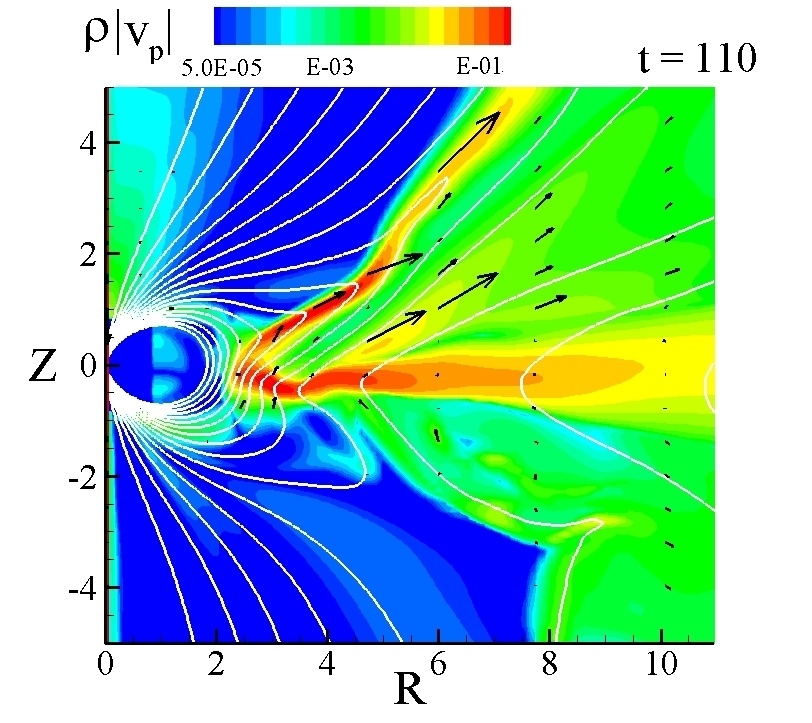}
        \caption{{\it Pure Stellar Dipole:} Poloidal mass flux $\rho |{\bf v}|$ (color), magnetic 
field lines (white lines) and coronal poloidal mass flux vectors $\rho {\bf v}_p$ at $t=110$ for the case of a pure stellar dipole field. Note the presence of two outflows, a magnetospheric wind and a disc wind.} 
\label{fig:pure_t=110}
\end{figure}

This case is useful for comparison with Lovelace et al. (2010) and for future runs that include a large scale disc-field.   Figure \ref{fig:pure_star+out} shows that
the outflows are quasi-episodic and dominantly asymmetric.
      Two types of outflows are observed.
        In the first, mass accretes onto the star via a funnel flow onto either the top or the bottom of the star. 
        At the same time, in the opposite hemisphere, the dipole loop expands and reconnects forming magnetic islands. 
         These magnetic islands carry off matter. 
         These were the types of outflows seen by Hayashi et al. (1996) for instance, but without the funnel flow, since their simulations were not bipolar. As in Hayashi we see a current sheet in the middle of the expanding loop. However, we do not see any significant heating of the gas as the loop expands. These outflows tend to have a greater velocity. In the second type of outflow, the disc pushes on the dipole, but has insufficient pressure to accrete onto the star. The magnetosphere re-expands, generating an episodic outflow (see Fig. \ref{fig:pure_t=110}). Our results are in close agreement with Lovelace et al (2010) that found episodic outflows for stellar dipoles and Lii et al. (2014) who found episodic outflows for stellar dipoles in the strong propeller regime. Both types of outflows occur on time scales $T \sim 10 \sim 0.35 y$ with a time scale $T \sim 50 \sim 17 y$ between events.     

\begin{figure}
                \centering
                \includegraphics[width=0.5\textwidth]{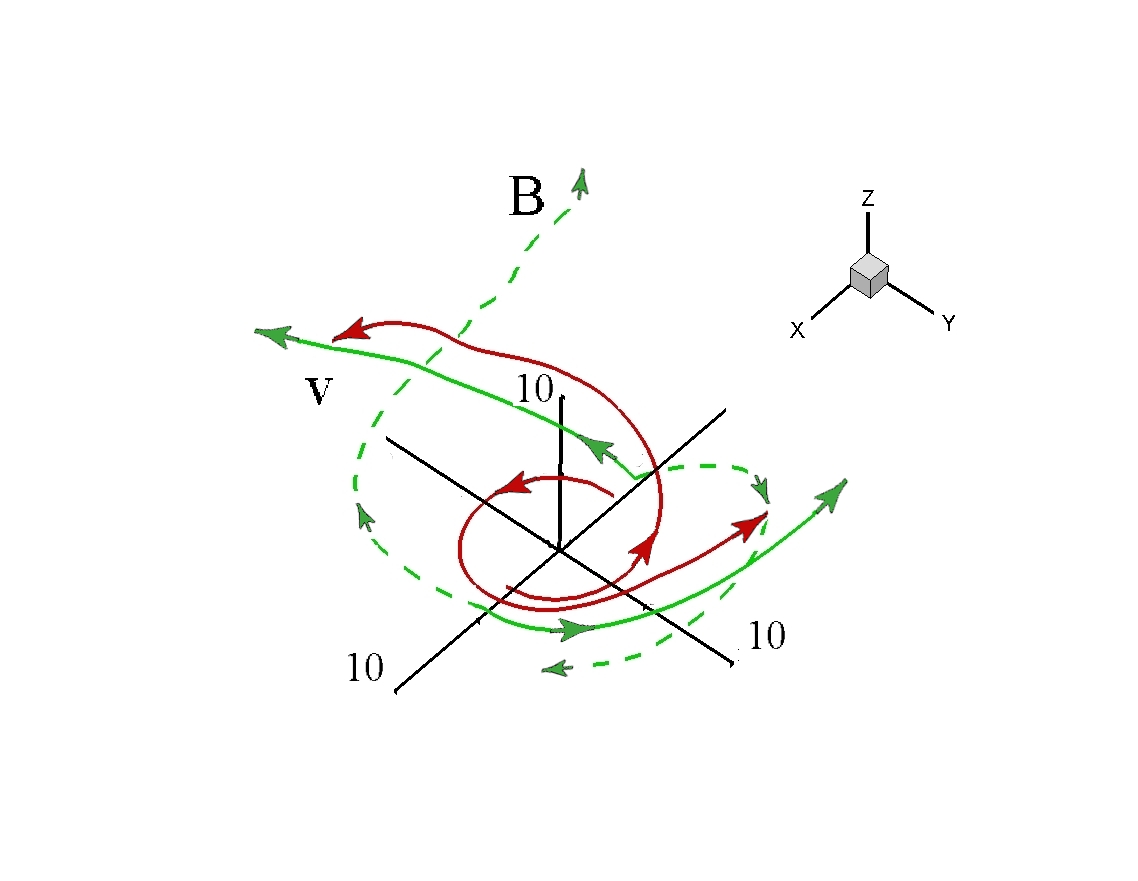}
        \caption{{\it Pure Stellar Dipole:} The solid red lines are streamline of the magnetospheric wind while  the solid green lines are
the streamlines for the disc wind.
  The dashed green lines are magnetic field lines anchored at the same point in the disc.  All data is at $t=110$ for 
a pure stellar dipole field with
 $\mu = 30$.} 
\label{fig:Helix}
\end{figure} 

Two qualitatively different types of outflows are observed: magnetospheric winds and disc winds.   
    Figure \ref{fig:pure_t=110} shows a time when  both types of outflows occur simulaneously. 
      The magnetospheric winds are launched perpendicular to the field lines of the stellar dipole. In contrast, the disc wind is launched along open field lines piercing the disc and extending off to infinity. Both outflows are inherently asymmetric, collimated by a toroidal magnetic field begining in the disc and extending into the corona. The disc wind is driven by field lines that are open in one hemisphere but closed and connected to the star in the other hemisphere. This corresponds to an outflow event that was not correlated to any stellar accretion.  

    Figure  \ref{fig:Helix} shows sample streamlines and field
lines of the magnetospheric and disc winds along the length of the outflow.
  We note that the magnetospheric wind has a density $10$ times greater than the disc wind, but comparable velocity at fixed distance from the launching point. This may be useful in explaining observations where jet/counter jet densities differ by a factor of a few.
       Also we note that the toroidal magnetic field drops to zero in the center of the magnetospheric wind, indicating the presence of a 
neutral layer.

\subsubsection{Stellar Dipole and Parallel Disc-Field PDIP}

\begin{figure}
                \centering
                \includegraphics[width=.5\textwidth]{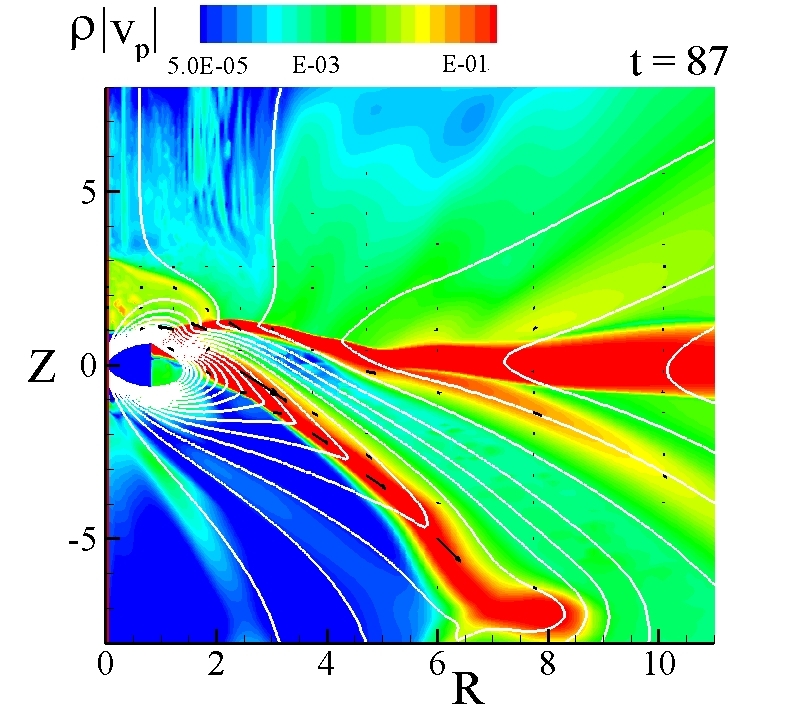}
        \caption{{\it Stellar Dipole plus Parallel Disc Field:}  Poloidal mass flux $\rho |\mathbf{v}|$ (color), magnetic field lines (white lines) and coronal poloidal mass flux vectors $\rho \mathbf{v}_p$ 
 at $t = 87$ for the case of a stellar dipole and a parallel disc field. Note the presence of a magnetospheric and disc wind, though the latter is considerably weaker.} 
\label{fig:t=87.jpg}
\end{figure}

\begin{figure}
                \centering
                \includegraphics[width=.5\textwidth]{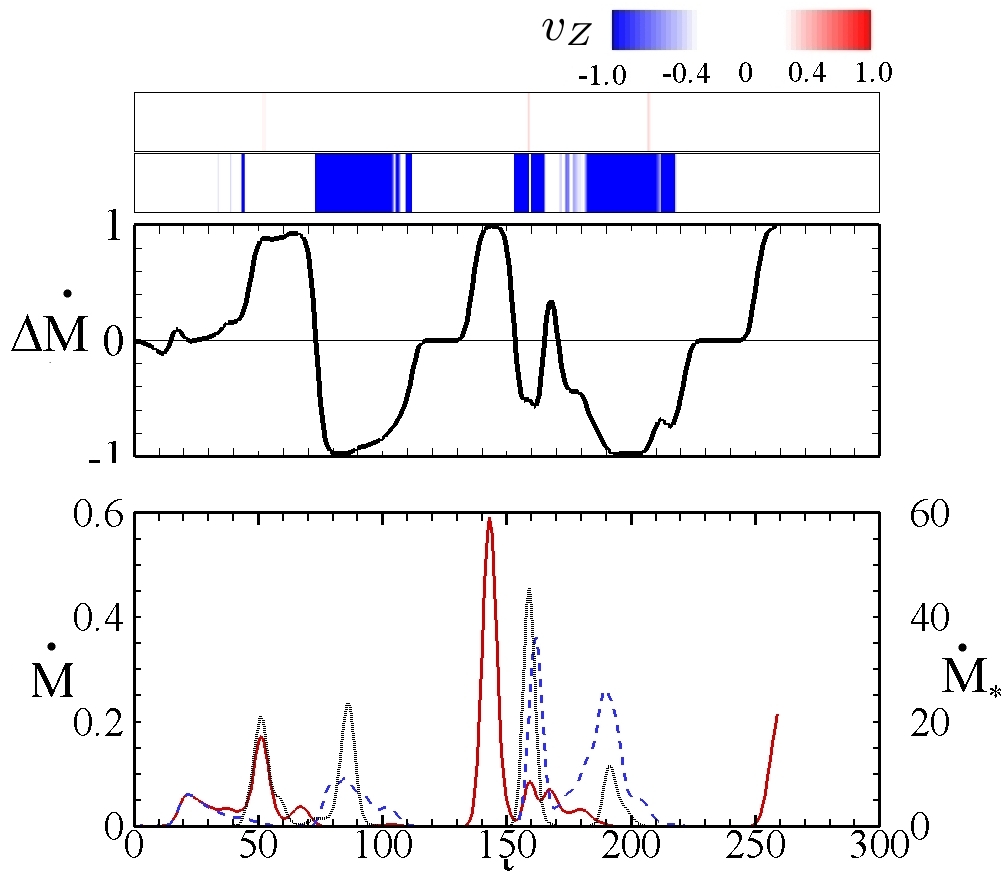}
        \caption{{\it Stellar Dipole plus Parallel Disc-Field:}
        Lower panel shows stellar mass accretion rate $\dot{M}_*$ (dotted black line line) and mass outflow $\dot{M}$ (red solid line for upper hemisphere; blue dashed line
 for lower hemisphere) as a function of time. The middle panel shows the mass flux difference, normalized to average total mass flux $\Delta \dot{M} = (\dot{M}_t - \dot{M}_b)/(\dot{M}_t + \dot{M}_b) $ as a function of time. The upper plot shows the average velocity of outflows, weighted by mass along the upper (red) and lower (blue) boundaries.
       The right-hand vertical axis for the outflows has been scaled so $\dot{M}/\dot{M}_* = 10^{-2}$.
   Note that there are  two types of outflows. 
   When the stellar accretion rate correlates with the matter outflow this corresponds to funnel flow accretion in one hemisphere and outflow in the other. These outflows have a greater average velocity. 
     Outflow when there is only small accretion corresponds to matter pushing against the magnetosphere in one hemisphere and being ejected outwards without matter accreting.} 
\label{fig:anti_star+out}
\end{figure}

These simulations feature two types of episodic outflows. 
     In the first, matter builds up at the magnetospheric boundary and begins to diffuse across the stellar dipole field.
     The initial top-bottom symmetry breaks spontaneously with
 the disc accreting  onto the star in a funnel flow in one hemisphere.
     In the opposite hemisphere,  matter is then pushed out by the re-expanding magnetosphere owing to the decrease in pressure from the disc due to the stellar accretion. 
     In this type of ejection the stellar accretion rate and the outflow rates correlate 
strongly.  Figure \ref{fig:t=87.jpg} shows a snapshot of an episodic
magnetospheric and disc wind outflow.
        Figure  \ref{fig:anti_star+out} shows the sporadic nature
 of these outflows which occur in this case  at $t=50,~85,~159,~\&~191$.
        
    The observed outflows are dominated by a magnetospheric wind, launched by a re-expanding magnetosphere, that had been compressed due to ram pressure from the disc. 
     In contrast to the pure dipole the disc wind component is suppressed. 
     This is a result of the field line geometry near the star. Owing to its centrifugal launching, the disc wind is dominated by outflows coming from the innermost region of the disc. 
        In the case of a pure stellar dipole field, the inner disc supports field lines
tilted away from the $Z-$axis which favors the formation of a disc wind.
        However, in the case of a stellar dipole and a parallel disc-field, the field lines in the inner disc are nearly vertical. The inclined field lines required for a disc wind are pushed out to larger radii and the strength of the wind is correspondingly decreased. 
         Interestingly, turning down the strength of the disc-field relative to the dipole 
field can turn  the inner disc wind back on by allowing for a field line geometry more favorable to outflows, as in the case of the pure dipole. 

      The parallel disc-field acts to increase the magnetospheric wind outflows by increasing the stellar mass flux. 
         The loss of angular momentum to outflows acts to increase the
  accretion rate of the disc (Dyda et al. 2012). 
         As a result,  accretion is  increased by roughly a factor of $4$, 
even though  the midplane magnetic $\beta$ at the magnetospheric boundary is roughly contant for the two cases. 
         This increased stellar $\dt{M}_*$ results, on  average in an increased 
 magnetospheric wind mass flux. 
          However, the effect is not great enough to vary the time scales of the episodic outflows which in both cases remain at roughly order $T \sim 50 \sim 17 y$. 
          This result suggests that the addition of a large scale disc-field will increase the overall mass flux of outflows, and increase the spread in the outflow's velocity distribution, owing to the fact that the magnetosphere is further compressed.

We can  turn off the magnetospheric wind and turn back on the inner disc wind by decreasing the strength of the dipole field. 
   We consider dimensionless dipoles $\mu = 10, ~20, ~30$. 
       At $t=0$ these correspond to $R_m = 2.8,~ 4.1,~ 5$, respectively. 
       All cases produce strong episodic outflows. 
       During such outflows we measure the magnetospheric radius and find $R_m = 1.4, ~2.1, ~2.2$. 
        In the case  $\mu = 10$ we are no longer in the propeller regime, that is to say $R_m \lesssim R_{\rm cr}$. 
        We can compare this to the case of the pure dipole where $R_m = 2.6, ~3.0, ~3.3$ during outflows. 
        In the case $\mu = 10$, the pure dipole outflows are stronger than
  those with a parallel disc-field by a factor of $\sim 2-3$. 
  In contrast, the $\mu = 20, 30$ parallel disc-field cases are stronger than the pure dipole cases by a factor of $\sim 2-5$. 
    This points up the importance of the field line geometry in determining 
the strength of outflows.
     Adding magnetic flux in a disc field can hinder outflows if it compresses the magnetospheric radius to values less than $R_{\rm cr}$.
     On the other hand, if $R_m > R_{\rm cr}$ then adding a large scale field can enhance outflows by increasing the accretion rate and providing more energy to drive outflows. 

\begin{table}
\begin{center}
  \begin{tabular}{ | c  l  c  c  c  c  c|}
                                                                 \\\hline
$\mu$   & Type    & $R_m$  &  $\dt{M}$ & $\dt{M}/\dt{M}_* (\%)$ &   v    &   $v/v_K(R_m)$     \\ \hline \hline
10      & DIP    &  2.6   &   0.055   &      1.7               &  0.48  &       0.77             \\
        & PDIP    &  1.4   &   0.009   &      0.4               &  0.20  &       0.40             \\
20      & DIP    &  3.0   &   0.078   &      0.7               &  0.49  &       0.85             \\
        & PDIP    &  1.8   &   0.141   &      4.2               &  0.23  &       0.31             \\
30      & DIP    &  3.3   &   0.083   &      0.1               &  0.33  &       0.60             \\
        & PDIP    &  2.1   &   0.364   &      0.8               &  0.38  &       0.55             \\
        \hline \hline
  \end{tabular}
\end{center}
\caption{Magnetospheric radius $R_m$, mass flux $\dot{M}$, characteristic velocity $v$ and scaled velocity $v/v_K(R_m)$ for different stellar dipoles $\mu$ in the case of a pure stellar dipole DIP and stellar dipole and parallel disc-field PDIP.}
\label{table:star_corr} 
\end{table}

Table \ref{table:star_corr} gives the magnetospheric radius $R_m$, mass outflow rate $\dt{M}$, characteristic outflow velocity $v$ and scaled outflow velocity $v/v_K(R_m)$ for different stellar dipole moments $\mu$ in the case of a pure dipole and disc-field during typical outflow events correlated with stellar accretion. We measure the characteristic outflow velocity by time averaging the velocity of the outflow at Z = 5 during the outflow event. 
     All outflows are magnetospheric winds, except for the anti-parallel  $\mu = 10$ case where magnetospheric ejections were absent
 because  $R_m < R_{\rm cr}$. 
    We notice that the magnetospheric radius is an increasing function of the dipole strength. For constant dipole strength, the addition of a disc-field further reduces $R_m$. 
         Further we note that outflow mass flux is typically a few percent of the  accretion mass flux. 
         The outflow velocity is roughly $75 \%$ of the Keplerian velocity at the launching radius. 
          We note that in the PDIP case $\mu=10$, $R_m < R_{ \rm{cr} }$. We are no longer in the propellor regime and this results in a lower outflow velocity. Increasing $\mu$ results in increasing the magnetospheric radius which in turn increases the strength of the propellor and the outflow velocity.   
         For pure stellar dipoles, the magnetospheric radius does not vary appreciably. The dipole is compressed until there is an equilibrium between the matter and magnetic pressure. 
          This leads to little variation in the size of typical outflows. 
          We note that in the case of $\mu = 10$, despite the fact that propeller has been turned off, the presence of a dipole field does inhibit accretion on the star. 
          This induces disc oscillations which leads to asymmetries in the outflows. This effect persists until late times, at which point the magnetosphere has been completely compressed by the advected large scale field and outflows become symmetric. 

The largest outbursts were not correlated to mass accreting on the star. 
      Rather there was a re-expantion of the magnetosphere, launching a ``knot" of matter. 
      The characteristic density of the ``knot" was $\rho \sim 20$, roughly 10 times greater than magnetospheric ejection for the pure dipole. 
      This accounts for the mass flux being a factor of $10$ greater. 
      We note the velocity of these much stronger outflows are roughly a factor of $2$ slower than the corresponding magnetespheric ejections in the pure dipole case.

\subsubsection{Dipole and Anti-Parallel Disc-Field ADIP}

    In the case of anti-parallel dipole and disc fields, reconnection occurs as the disc-field is advected inwards and runs into the stellar dipole field.   
    We find funnel flow accretion to the poles of the star as found earlier by Miller and Stone (1997).
     This type of accretion is favored for this field geometry because at the X-point there is no magnetic pressure and matter can move freely onto the field lines connecting the disc and the poles of the star.

As magnetic flux from the disc-field advects inwards, the X-point in the disc moves outwards. 
     This has the effect of moving the open field lines threading the disc outwards. 
      This causes the characteristic outflow speed, set by the local Keplerian velocity, to decrease. 
        Outflows are hence diminished in this type of field configuration, and should die down at late times as sufficient magnetic flux is advected inwards and the inner disc field lines move outward.  

\begin{figure}
                \centering
                \includegraphics[width=.5\textwidth]{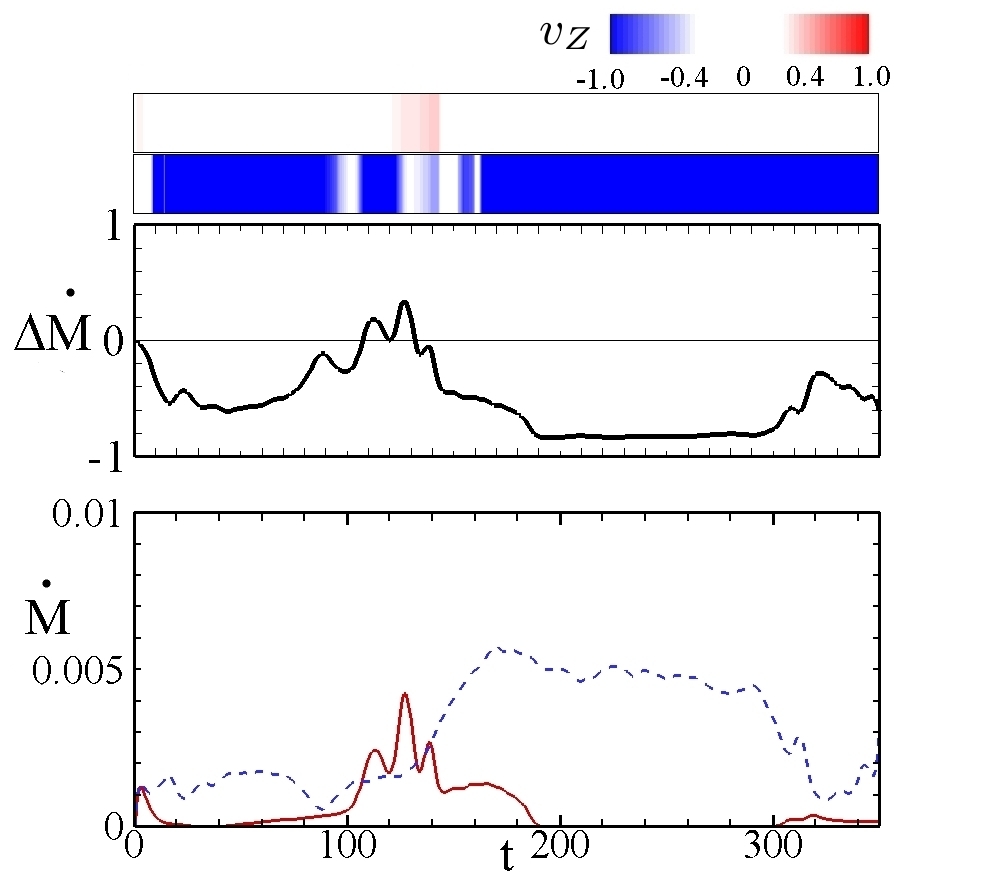}
        \caption{{\it Stellar Dipole plus Anti-Parallel Disc-Field:}
     The lower panel shows the mass flux $\dot{M}$ through the upper (solid black line) and lower (dashed red line) boundaries as a function of time.
       The upper panel shows the  mass flux difference, normalized to average total mass flux $\Delta \dot{M} = (\dot{M}_t - \dot{M}_b)/(\dot{M}_t + \dot{M}_b) $ as a function of time. We note that outflows are $\sim 100$ times weaker than in the DIP and PDIP cases.} 
\label{fig:mdot_anti}
\end{figure}

\begin{figure}
                \centering
                \includegraphics[width=.5\textwidth]{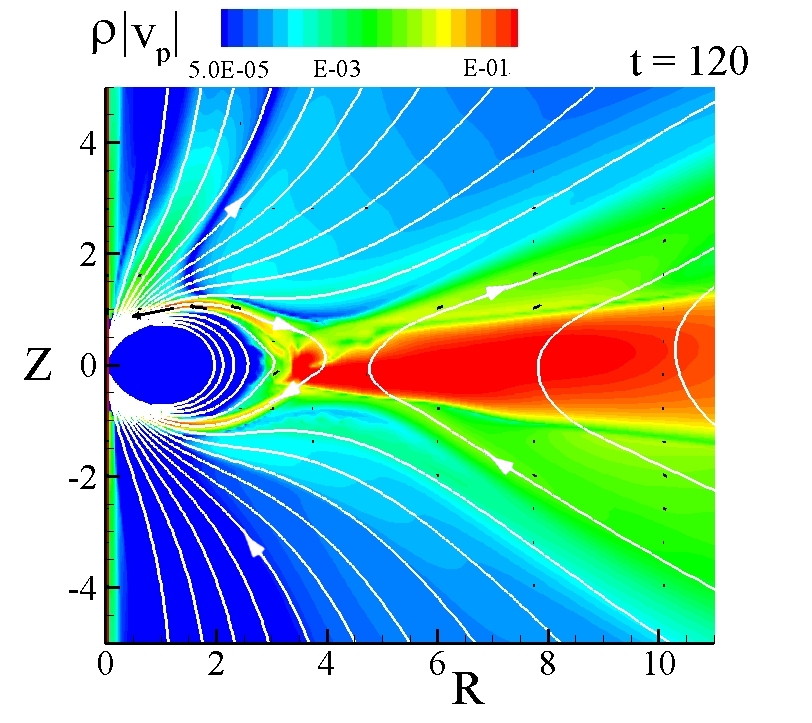}
        \caption{{\it Stellar Dipole plus Anti-Parallel Disc Field ADIP:}  Poloidal mass flux $\rho |\mathbf{v}|$ (color), magnetic field lines (white lines) and coronal poloidal mass flux vectors $\rho \mathbf{v}_p$ at $t = 120$ for the case of a stellar dipole and an anti-parallel disc field. We note the weakeness of the disc wind despite the funnel flow accretion onto the star.} 
\label{fig:120.jpg}
\end{figure}

     Figure \ref{fig:mdot_anti} shows the mass flux $\dt{M}$ through the upper and lower hemispheres.
     As the large scale field is advected inwards and annihilates with the dipole, the X-point moves into the lower hemisphere. 
     This causes a top/bottom disc asymmetry, as matter preferentially falls into the lower hemisphere. 
        This leads to a stronger outflow rate in the lower hemisphere at late times.
   Figure \ref{fig:120.jpg} shows a snapshot of the disc and poloidal magnetic
   field

\begin{figure}
                \centering
                \includegraphics[width=.5\textwidth]{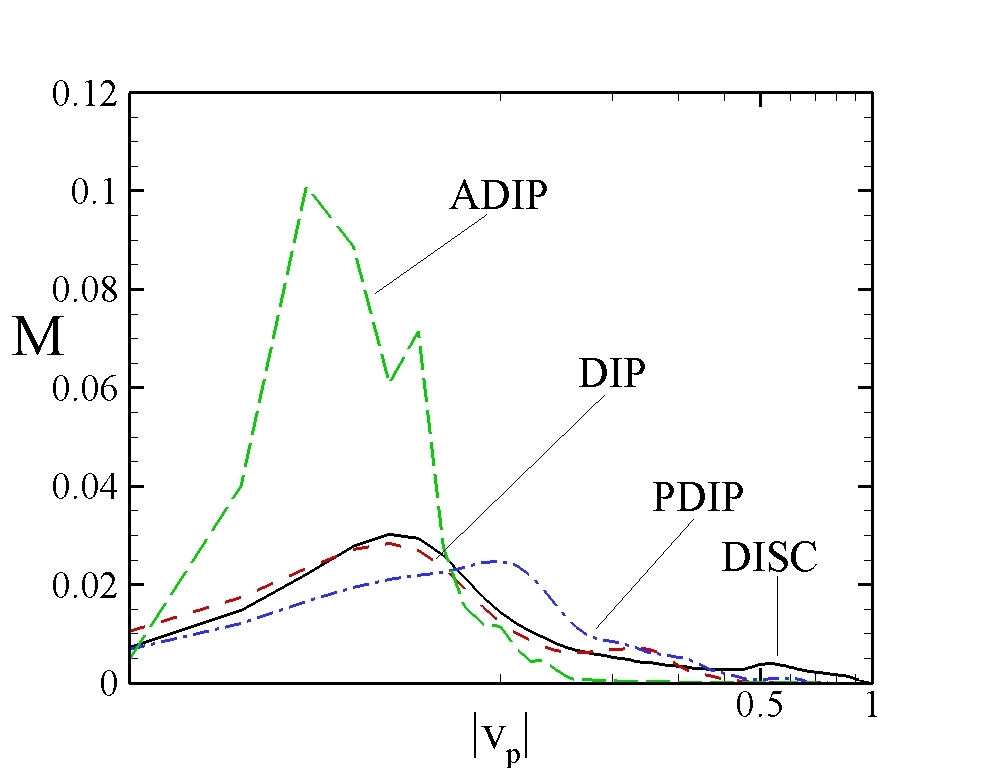}
        \caption{Fraction of total mass outflowed as a function of poloidal velocity for DISC (solid black line), DIP (red dashed line) , PDIP (blue dash-dot-dot line) and ADIP (green long dash line) simulations.
    The velocity cutoff is set by the Keplerian velocity scale at the innermost launching radius. In the case of the disc-field, this is at
the surface of the star and therefore $v_{max} = 1$. In cases with a dipole field this is the magnetospheric radius $R_m > 1$ and therefore $v_{\rm max} < 1$. For the PDIP case, the disc field compresses the magnetosphere and the peak appears at larger $v_p$ than for the DIP case. We show the plot for $v_p > 0$ since the full distribution is symmetric.  
    } 
\label{fig:histogram}
\end{figure}

   The field line geometry is not favorable for launching inner disc winds. Despite initially being in contact with the disc (see Fig. \ref{fig:dipole_ic}), field lines anchored to the star quickly move into the corona 
   so that there is no efficient mechanism for mass loading onto them
 from the disc.
   The innermost open field lines anchored in the disc are at $R > 7$, so that the disc wind is  weak. 
   Based on the velocity cutoff of $v_{max} \approx 0.2 $ in Fig. \ref{fig:histogram} we would have estimated this launching radius as $R \approx 25$. 

The stellar mass accretion rate $\dt{M}_* \sim 0.5$ which is a factor of $\sim 10$ times smaller than the accretion rate of the pure disc-field.  
     In contrast with the pure dipole, annihilations of the disc-field with the stellar dipole reduces the dipole strength and allows for matter to more easily  accrete onto the star. 
      This leads to a nearly constant accretion rate, unlike the case of the pure dipole where accretion occurs in episodic funnel flow events.  
      The total mass accretion during the interval $0 < t < 200$ is comparable in the cases of the pure dipole and the dipole with parallel disc-field. 
      This suggests that the total energy loss in each case is the same. 
      The parallel configuration, owing to the lack of outflows, converts a much smaller fraction of this energy into outflows.

Figure \ref{fig:histogram} shows the fraction of mass in outflows with a given poloidal velocity for the various combinations of stellar dipole and disc field. 
       Over such long time scales the mass outflow is to a high degree
  of accuracy symmetric between the two hemispheres so we only plot our results for $v_p > 0$. 
     We further note several interesting features. 
        The distributions have a sharp velocity cutoff at $v_{max} = 0.5, ~0.65, ~1.0$ in the cases of the DIP, PDIP and DISC respectively. 
    These velocities correspond to those
  of a Keplerian disc at radii $R_{K} = 1.6,~ 1.33, ~1.0$. 
    The velocity cutoff directly corresponds to the Keplerian velocity at the innermost launching radius. 
    In the case of the disc-field, this is the surface of the star and therefore $v_{max} = 1$. 
        In cases with a dipole field this is the magnetospheric radius $R_m > 1$ which leads to $v_{\rm max} < 1$. 
         In the case of the parallel disc-field, the  disc-field compresses the dipole field and decreases the magnetospheric radius. 
         Hence the maximum observed velocity is higher than in the case of a pure dipole. Also the peak in the distribution is shifted to higher velocities.
         This provides information about the field structure near the star: 
          A large scale field will have outflows over the entire velocity range whereas the presence of a dipole may have a cutoff at $v_{\rm max} 
  \propto R_{m}^{-3/2}$ since outflows will not be launched at radii $R < R_m$.

\section{Conclusion}
Here we summarize some of our main conclusions.
\smallskip

\noindent{\bf 1.} The distributed disc-fields (DISC) in the
absence of a stellar field produce symmetric outflows for low  density, low
midplane-$\beta$ discs.  
      This limit agrees with study by Fendt and Sheikhnezami  (2013).
For larger disc densities,  there is a   decoupling of Alfv\'en waves between the upper and lower hemispheres and this leads to oscillations in the outflows between the two hemispheres.  
     For even higher disc densities there the jet outflows
are strongly asymmetric with the outflow changing episodically from one
hemisphere to the other.
\smallskip

\noindent{\bf 2.} For a stellar dipole field in the absence of a disc-field (DIP),  two types of outflows are observed - magnetospheric winds from the magnetospheric
radius $R_m$ and disc winds from somewhat larger radii.
   Both components have similar velocities but the magnetospheric winds 
have significantly higher densities and thus carry most of the mass outflow
$\dt{M}$.   
    The asymmetry of the outflows between the top and bottom
hemispheres increases as the magnetospheric radius $R_m$ increases beyond the corotation radius $R_{\rm cr}$ as found earlier by Lovelace  et al.
(2010).
\smallskip

\noindent{\bf 3.} Introducing a disc-field parallel to the
stellar dipole field (PDIP)  acts to reduce the magnetospheric radius $R_m$ and changes the inner disc field  structure. 
    The field lines in the inner disc field lines are more aligned
with the $Z-$axis and as a result the disc wind component is
suppressed.
    At the same time  the characteristic outflow velocity of the magnetospheric winds is increased owing to the reduced magnetospheric radius. Outflows are correspondingly more powerful.  
     The outflows are episodic and strongly asymmetric about the
equatorial plane.  The asymmetry of the outflows is not
 affected  by
the initial midplane plasma $\beta$ of the disc.
    \smallskip

\noindent{\bf 4.} For a disc-field which is anti-parallel to the stellar dipole field (ADIP), we observe
reconnection between the field components analogous in some respects to the reconnection between the subsonic solar wind magnetic field and the earth's
dipole field at the magnetopause which is inside the Earth's bow shock (e.g., Hargreaves 1992).  
    This reconnection acts to prevent a build up of poloidal magnetic flux outside the magnetosphere.  As a result magnetospheric outflows are
suppressed.   At the same time accretion  to the star is enhanced. 

\smallskip

  Our results suggest that multi-component outflows, disc winds and
  magnetospheric winds, occur for a stellar dipole field in the absence
  of a large-scale disc-field.
      Inward advection of a large scale disc-field acts to 
 turn off the disc wind component so that there is only a
 magnetospheric wind.
     We observe a cutoff in the maximum velocity of outflows related to the inner launching radius. This suggests that by measuring the velocity of outflows, one can estimate the launching radius. Measurement of the dipole component of the magnetic field would then allow one to estimate the matter pressure at the magnetospheric boundary. 

Ellerbroek et al. (2013) found they could reproduce the knotted structure of outflows from HH1042 with a simple periodic outflow model. One possible solution that was consistent with observations was one where the jet and counter jet were out of phase by a half period. Observations were consistent with a period $T \sim 10-100 \rm{y}$ Our DISC simulation with intermediate $\beta$ enters into a regime with period $T = 2.82 y \left( R_0 / 0.1 \rm{AU}\right)^{3/2}$. This agrees with the observed periods with a reasonable tuning of the inner disc length scale $R_0$.

Woitas (2002), Hartigan \& Hillenbrand (2009) and Melnikov et al. (2009) have found that RW Aur A has a jet with velocity roughly 2 times greater and density roughly 2-3 times as large as its counter jet. This is the type of behaviour seen in our intermediuate $\beta$ DISC run where in the oscillatory phase, the bottom jet mass flux peaks at a rate roughly 3 times greater than the top jet and has a velocity 2-3 times greater.

\section*{Acknowledgments}

Resources supporting this work were provided by the NASA High-End
Computing (HEC) Program through the NASA Advanced Supercomputing
(NAS) Division at the NASA Ames Research Center and the NASA
Center for Computational Sciences (NCCS) at Goddard Space Flight
Center. The research was supported in part by NASA grant NNX14AP30G,
and NSF grant AST-1211318. We thank the anonymous referee for their very thorough
review of our work.


\label{lastpage}

\end{document}